\documentclass[aps,pre,subeqn,nofootinbib,amsmath,amsfonts]{revtex4}
\usepackage{graphicx, xcolor}
\usepackage{longtable}
\usepackage{subfigure}
\usepackage{euscript}
\usepackage[hyperfootnotes=false]{hyperref}

\newcommand{\eqr}[1]{Eq.~\eqref{#1}}

\newcommand{\secr}[1]{Sec.~[\ref{#1}]}
\newcommand{\figr}[1]{Fig.~[\ref{#1}]}
\newcommand{\figsr}[2]{Fig.~[\ref{#1}-\ref{#2}]}

\newcommand{\tblr}[1]{Table~[\ref{#1}]}
\newcommand{\tblsr}[2]{Tables~[\ref{#1}-\ref{#2}]}
\newcommand{\appr}[1]{Appendix~[\ref{#1}]}

\newcommand{\velocity}{{\rm{v}}}
\newcommand{\vvelocity}{{\bf{v}}}
\newcommand{\vJ}{{\textbf{J}}}
\newcommand{\vR}{{\mathbf{r}}}
\newcommand{\vRt}{\vR,\,t}
\newcommand{\vg}{{\textbf{g}}}

\newcommand{\spd}{\!\cdot\!}
\newcommand{\PCDF}[2]{{\frac{\partial #1}{\partial x_{#2}}}} 

\newcommand{\vn}{{\mathbf{n}}}

\newcommand{\vq}{\mathbf{q}}

\newcommand{\xs}{x^{\scriptstyle s}}

\newcommand{\xgsb}{x^{g,s}}
\newcommand{\xlsb}{x^{\ell,s}}
\newcommand{\xg}{x^{g}}
\newcommand{\xl}{x^{\ell}}

\newcommand{\rs}{\vR^{\scriptstyle s}}              

\newcommand{\profilescale}{0.4}

\makeatletter%
\renewcommand*\env@matrix[1][c]{\hskip -\arraycolsep%
  \let\@ifnextchar\new@ifnextchar%
  \array{*\c@MaxMatrixCols #1}}%
\makeatother%

\graphicspath{{Figures/}}

\begin{document}
\title{Transfer coefficients for the Gibbs surface in a two-phase mixture in the non-equilibrium square gradient model.}
\author{K.~S.~Glavatskiy$^{1}$}
\author{D.~Bedeaux$^{1,2}$}
\affiliation
{%
$^{1}$Department of Chemistry, Norwegian University of Science and Technology, NO 7491 Trondheim, Norway.\\
$^{2}$Department of Process and Energy, Technical University of Delft, Leeghwaterstr 44, 2628 CA Delft, The Netherlands.
}%
\date\today
\begin{abstract}
In this paper we calculate the transfer coefficients for evaporation and condensation of mixtures. We use the continuous profiles of various thermodynamic quantities through the interface,
obtained in our previous works using the square gradient model. Furthermore we introduce the Gibbs surface and obtain the excess entropy production for a surface. Following the traditional
non-equilibrium thermodynamic approach we introduce the surface transfer coefficients which we are able to determine from the continuous solution. The knowledge of these coefficients is
important for many industrial applications which involve transport through a surface, such as for instance distillation. In our approach the values of the local resistivities in the liquid and
the vapor phases are chosen on the basis of experimental values. In the interfacial region there are small peaks in these resistivities. Three amplitudes control the magnitude of these peaks.
Possible values of these amplitudes are found by matching the diagonal transfer coefficients to values predicted by kinetic theory. Using these amplitudes we find that the value of the cross
resistivities is 1-2 orders of magnitude higher then the one from kinetic theory. The results of both kinetic theory and molecular dynamics simulations support the existence of small peaks in
the local resistivities in the interfacial region. The square gradient approach gives an independent way to determine the transfer coefficients for surfaces. The results indicate that kinetic
theory underestimates the interfacial transfer coefficients in real fluids.
\end{abstract}
\maketitle
%
%
\numberwithin{equation}{section}
\section{Introduction}

In two earlier articles \cite{glav/grad1, glav/grad2} we developed the general approach for the square gradient description of an interface between two phases in non-equilibrium $n$-component
mixtures. Using that approach it is possible to determine the continuous profiles of all variables through the interface during, for instance, evaporation and condensation. In this paper we will
use these results to obtain the transfer coefficients for heat and mass transfer through the liquid vapor interface. The values of these transfer coefficients, or even their order of magnitude,
is extremely important for industrial processes which involve evaporation and/or condensation of mixtures. Among these processes is, for instance, distillation, when one needs to separate
components with different volatilities. As this involves evaporation and/or condensation repeatedly many times, it is very important to know the exact effect of the surface. Some values of the
interfacial transfer coefficients may favor transport of one component, while the others may favor adsorption of a component at the surface. Of particular interest are the values of the cross
coefficients, which contribute to reversible transport, and which are in most descriptions neglected \cite{kjelstrupbedeaux/heterogeneous}.

A number of different methods have been used to obtain the surface transfer coefficients for one-component systems: experiments \cite{Fang1999a, Bedeaux1999b, Mills2002, James2006}, molecular
dynamic simulations \cite{Rosjorde2000, Rosjorde2001, Kjelstrup2002, Simon2004, jialin/longrange} , kinetic theory \cite{Pao1971a, Sone1973, Cipolla1974, Bedeaux1990}. In a paper coauthored by
one of us \cite{bedeaux/vdW/III} the interfacial transfer coefficients obtained from the gradient theory for a one-component system were calculated and compared to the data in the above
references. Even for one-component systems the database of interfacial transfer coefficients is poor and these data are pretty scattered. The situation is even worse for mixtures. There are only
few experiments available for several systems \cite{James2006, Mills2002} at a very restrictive range of conditions, i.e. for instance, at infinite dilution. No molecular dynamic simulations are
available yet. The only source of the values of interfacial coefficients is kinetic theory \cite{Cipolla1974, Bedeaux1990}. This theory is most appropriate for short range potentials and low
density gases. There is evidence from molecular dynamic simulations for longer range potentials \cite{jialin/longrange} that the coupling transfer resistivities for liquid-vapor interfaces of
real fluids are substantially larger than those predicted by kinetic theory.

In the first paper \cite{glav/grad1} we discussed the balance equations and the Gibbs relation for the square gradient model of mixtures. The Gibbs relation enabled us to derive the entropy
production. It followed that if one uses as thermodynamic forces the gradient of the inverse temperature, $\mathbf{\nabla }(1/T)$, minus the gradients of the chemical potential differences with
the $n$th component divided by the temperature, $-\mathbf{\nabla}[(\mu _{j}-\mu _{n})/T]$, for $j=1,...,n-1$, and minus the gradient of the velocity gradient divided by the temperature,
$-(\mathbf{\nabla} \vvelocity)/T$, that the conjugate fluxes are the total heat flux, $\mathbf{J}_{q}$, the diffusion fluxes of the first $n-1$ components relative to the barycentric frame of
reference, $\mathbf{J}_{j}$, and the viscous pressure tensor. Linear laws relating these forces and fluxes could then be given. Together with the balance equations it is then possible to
calculate the profiles of all the variables. In the second paper \cite{glav/grad2} we gave details of the numerical solution procedure for stationary states. We defined excess densities for an
arbitrary dividing surface and verified that the non-equilibrium surface, as described by these excess densities following Gibbs, is in local equilibrium. Our work \cite{glav/grad1, glav/grad2}
extended the analysis given by Johannessen et al. \cite{bedeaux/vdW/I, bedeaux/vdW/II}.

Given the validity of local equilibrium for the description in terms of the Gibbs excess densities, it is possible to develop a description using non-equilibrium thermodynamics as explained in
the monograph by Kjelstrup and Bedeaux \cite{kjelstrupbedeaux/heterogeneous}. This is much easier than the continuous description. As we will verify in this paper the expression for the excess
entropy production of a surface has the general form
\begin{equation}
\widehat{\sigma }_{s}=\sum_{i}{J_{i}X_{i}}  \label{eq/Introduction/01}
\end{equation}%
In this expression $J_{i}$ are the heat and mass fluxes through the surface and $X_{i}$ are the jumps in the intensive variables across the interface. In non-equilibrium one uses the finite
jumps of the temperature and chemical potentials across the surface, which lead to a non-zero entropy productions in the interfacial region. These jumps become the driving forces for the heat
and mass transport through the interface. Following the traditional approach of non-equilibrium thermodynamics we then write the linear force-flux relations. These expressions use the
interfacial resistivities or transfer coefficients which are the key interest of this paper.

Having the continuous profiles of thermodynamic quantities obtained from the non-equilibrium gradient model we are able to calculate these resistivities independently. This gives a way to
determine the coefficients and therefore a possible source for comparison with future experiments and simulations.

In \secr{sec/Entropy} we derive the expression for the local entropy production for stationary states in the continuous description. In \secr{sec/Excess} we discuss the properties of the excess
quantities in three-dimensional space. We consider how the stationary state condition simplifies the non-equilibrium expressions. In \secr{sec/ExcessEntropy} we obtain the expression for the
excess entropy production in an interfacial region. In \secr{sec/TransportCoef} we give the force-flux relations and discuss the interfacial resistivity coefficients. We consider different sets
of coefficients which are associated with different variables: gas- and liquid- side coefficients, as well as mass and molar coefficients. Any coefficient of one set is determined by the
coefficients of the other set and equilibrium properties of pure bulk components. These sets are therefore equivalent. In \secr{sec/TwoComponent} we proceed to a particular two-component mixture
and specify further details. In \secr{sec/Methods} we discuss the different methods to obtain the resistivities from a non-equilibrium continuous solution. We give the results of our analysis in
\secr{sec/Results}. We analyze extensively different aspects of the problem and find the values of parameters, which make the interfacial resistivities obtained from the continuous solution of
the gradient model to match kinetic theory. We give a discussion and concluding remarks in \secr{sec/Discussion}.

\section{Local entropy production.}\label{sec/Entropy}
\subsection{Gibbs-Duhem equation.}\label{sec/Entropy/Gibbs-Duhem}

Consider a two phase $n$-component mixture. Let $T$ be the temperature field in this region, $\psi_{i} \equiv \mu_{i}-\mu_{n}$ be the chemical potential differences and $p$ be the scalar
pressure, which in case of a planar interface coincides with the parallel pressure $p_{\parallel}$. Furthermore, let $u$, $s$, $v$ be the mass specific internal energy, entropy and volume
respectively, $\rho \equiv 1/v$ be the overall mass density and $\xi_{i} \equiv \rho_{i}/\rho$ be the mass fraction of the $i$-th component. The explicit expressions for these quantities are
given in \cite{glav/grad1}. It was found that the Gibbs equation for such a two phase system is given by:
\begin{equation}\label{eq/Entropy/Gibbs/01}
T(\vRt)\,\frac{ds(\vRt)}{dt} = \frac{du(\vRt)}{dt} - \sum_{i=1}^{n-1}{\psi_{i}(\vRt)\frac{d\xi_{i}(\vRt)}{dt}} + p(\vRt) \frac{dv(\vRt)}{dt} - v(\vRt)\,\velocity_{\beta}(\vRt)\frac{\partial
\gamma_{\alpha\beta(\vRt)}}{\partial x_{\alpha}}
\end{equation}%
where $\vvelocity$ is the barycentric velocity, $d/dt$ is a substantial time derivative and we use the summation convention over double Greek indices. $\gamma_{\alpha\beta}$ is the tension
tensor, which is given by
\begin{equation}\label{eq/Entropy/Gibbs/02}
\gamma_{\alpha\beta}(\vRt) = \kappa\,{\PCDF{\rho(\vRt)}{\alpha}}{\PCDF{\rho(\vRt)}{\beta}} + {\sum\limits_{i=1}^{n-1}{\kappa_{i}\,\Big({\PCDF{\xi_{i}(\vRt)}{\alpha}}{\PCDF{\rho(\vRt)}{\beta}} +
{\PCDF{\rho(\vRt)}{\alpha}}{\PCDF{\xi_{i}(\vRt)}{\beta}}\Big)}} +
{\sum\limits_{i,j=1}^{n-1}{\kappa_{ij}\,{\PCDF{\xi_{i}(\vRt)}{\alpha}}{\PCDF{\xi_{j}(\vRt)}{\beta}}}} %
\end{equation}%
It is non-zero in particular in the interfacial region, where the gradient variables $\nabla\,\rho$ and $\nabla\xi_{i}$ are significant. All thermodynamic densities (except the entropy) have
gradient contributions. Foe explicit expressions we refer to \cite{glav/grad1}. These densities are related by
\begin{equation}\label{eq/Entropy/Gibbs/03}
u(\vRt) = \mu_{n}(\vRt) + \sum_{i=1}^{n-1}{\psi_{i}(\vRt)\,\xi_{i}(\vRt)} - p(\vRt)\,v(\vRt) + T(\vRt)\,s(\vRt)
\end{equation}%

Substituting \eqr{eq/Entropy/Gibbs/03} into \eqr{eq/Entropy/Gibbs/01} we obtain
\begin{equation}\label{eq/Entropy/Gibbs/04}
s\,\frac{dT}{dt} + \frac{d\mu_{n}}{dt} + \sum_{i=1}^{n-1}{\xi_{i}\frac{d\psi_{i}}{dt}} - v\,\frac{dp}{dt} - v\,\velocity_{\beta}\frac{\partial \gamma_{\alpha\beta}}{\partial x_{\alpha}} = 0
\end{equation}%
This is the Gibbs-Duhem equation for a two-phase multi-component mixture. Note that since $\psi_{i} \equiv \mu_{i}-\mu_{n}$, we have
\begin{equation}\label{eq/Entropy/Gibbs/05}
\frac{\partial \mu_{n}}{\partial x_{\beta}} + \sum_{i=1}^{n-1}{\xi_{i}\frac{\partial \psi_{i}}{\partial x_{\beta}}} = \sum_{i=1}^{n}{\xi_{i}\frac{\partial \mu_{i}}{\partial x_{\beta}}}
\end{equation}%
which is the usual contribution to the Gibbs-Duhem equation associated with the chemical potentials.

For a stationary state the derivative $\partial / \partial t = 0$ and \eqr{eq/Entropy/Gibbs/04} takes the following form
\begin{equation}\label{eq/Entropy/Gibbs/04a}
\velocity_{\beta}\,\left( s\,\frac{\partial T}{\partial x_{\beta}} + \frac{\partial \mu_{n}}{\partial x_{\beta}} + \sum_{i=1}^{n-1}{\xi_{i}\frac{\partial \psi_{i}}{\partial x_{\beta}}} -
v\,\frac{\partial \sigma_{\alpha\beta}}{\partial x_{\alpha}}\right) = 0
\end{equation}%
where $\sigma_{\alpha\beta} = p\,\delta_{\alpha\beta} + \gamma_{\alpha\beta}$ is the thermodynamic pressure tensor.

\subsection{Entropy balance.}\label{sec/Entropy/Balance}

The entropy balance equation is
\begin{equation}\label{eq/Entropy/Balance/01}
\rho\,\frac{ds}{dt} = - \nabla\spd\vJ_{s} +  \sigma_{s}
\end{equation}%
with the entropy flux $\vJ_{s} \equiv \vJ_{s,tot} - \rho s\vvelocity$ and the entropy production $\sigma_{s}$. These were found to be
\begin{subequations}\label{eq/Entropy/Balance/02}
\begin{equation}\label{eq/Entropy/Balance/02a}
\vJ_{s} = \displaystyle \frac{1}{T}\left(\vJ_{q} - \sum_{i=1}^{n-1}{\psi_{i}\,\vJ_{i}} \right)
\end{equation}
\begin{equation}\label{eq/Entropy/Balance/02b}
\sigma_{s} = \displaystyle \vJ_{q}\spd\nabla\frac{1}{T} - \sum_{i=1}^{n-1}{\vJ_{i}\spd\nabla\frac{\psi_{i}}{T}} - \pi_{\alpha\beta}\,\frac{1}{T}\frac{\partial \velocity_{\alpha}}{\partial
x_{\beta}}
\end{equation}%
\end{subequations}
where $\vJ_{q}$ and $\vJ_{i}$ are the heat and diffusion fluxes
\begin{equation}\label{eq/Entropy/Balance/03}
\begin{array}{rl}
\vJ_{q} &\equiv \displaystyle \vJ_{e} - \rho\,\vvelocity\,e - p\,\vvelocity - \pi\spd\vvelocity = \vJ_{e} - \vJ_{m}\,(e + p\,v) - \pi\spd\vvelocity
\\ \\
\vJ_{i} &\equiv \rho_{i}\,(\vvelocity_{i}-\vvelocity) = \vJ_{\xi_{i}} - \xi_{i}\,\vJ_{m}
\end{array}
\end{equation}%
where $\pi \equiv \pi_{\alpha\beta}$ is the viscous pressure tensor. The energy flux $\vJ_{e}$ and the mass fluxes $\vJ_{\xi_{i}} \equiv \rho_{i}\,\vvelocity_{i}$ and $\vJ_{m} \equiv
\rho\,\vvelocity$ are convenient quantities since in stationary states
\begin{equation}\label{eq/Entropy/Balance/03a}
\nabla\spd\vJ_{e} = 0 ,\quad \nabla\spd\vJ_{\xi_{i}} = 0 ,\quad \nabla\spd\vJ_{m} = 0
\end{equation}%
Furthermore, it follows from \eqr{eq/Entropy/Balance/01} that in stationary state
\begin{equation}\label{eq/Entropy/Balance/04}
\sigma_{s} = \nabla\spd\vJ_{s,tot} = \nabla\spd\vJ_{s} + \rho\,\vvelocity\spd\nabla s = \nabla\spd(\vJ_{s,tot} + \rho\,s\vvelocity)
\end{equation}%
Using \eqref{eq/Entropy/Gibbs/04a} and the conservation laws under stationary conditions, it is possible to show that
\begin{equation}\label{eq/Entropy/Balance/05}
\sigma_{s} = \vJ_{e}\spd\nabla\frac{1}{T} - \sum_{i=1}^{n}{\vJ_{\xi_{i}}\spd\nabla\frac{\mu_{i}}{T}} - \vJ_{m}\spd\nabla\frac{\velocity^{2}/2-\vg\spd\vR}{T} - \frac{\partial}{\partial
x_{\alpha}}\frac{\pi_{\alpha\beta}\velocity_{\beta}}{T}
\end{equation}%
The expression for the entropy production, used in \eqr{eq/Entropy/Balance/05} contains dependent fluxes $\vJ_{\xi_{i}}$ and $\vJ_{m}$ and thus force-flux relations cannot be obtained from it
directly.

\section{Excesses in three-dimensional space.}\label{sec/Excess}
\subsection{Definition of an excess.}\label{sec/Excess/Definition}

The definition of an excess requires the normal direction $\vn$ to be defined in the interfacial region. The surface may be curved and we may introduce curvilinear orthogonal coordinates
$(x_{1}, x_{2}, x_{3})$ with $\vR_{\perp} \equiv x_{1}$ being the normal coordinate and $\vR_{\parallel} \equiv (x_{2}, x_{3})$ being the tangential coordinates.

Let $\xgsb$ and $\xlsb$ be the boundaries of the interfacial region at the gas and liquid side respectively. Let $\phi$ be a function defined in the surface region. Furthermore let $\phi^{b}$,
where superscript $b$ stands either for $\ell$ or for $g$, be the function $\phi$ extrapolated from the bulk to the surface region. The extrapolation is done using the description in homogeneous
phases which does not contain gradient contributions. Outside of the interfacial region $\phi^{b}$ and $\phi$ are identical but inside the surface, $\phi^{b}$ in general differs from $\phi$. We
note the following identity for the extrapolated functions
\begin{equation}\label{eq/Excess/Definition/01}
\phi^{b}(x^{b,s}, \vR_{\parallel}) = \phi(x^{b,s}, \vR_{\parallel})
\end{equation}
Furthermore, for any function $F$
\begin{equation}\label{eq/Excess/Definition/02}
F^{b}(\ldots, \phi, \ldots) = F(\ldots, \phi^{b}, \ldots)
\end{equation}
in the interfacial region, since for the bulk functions outside the interfacial region it is identity. We note however, that even though \eqr{eq/Excess/Definition/01} and
\eqr{eq/Excess/Definition/02} are exact, any numerical procedure will break these equalities. This happens because the extrapolation procedure usually involves polynomials in order to fit an
actual curve, which introduces a non-zero error in the extrapolated curve.

We then can define the excess $\widehat{\phi}(\xs, \vR_{\parallel})$ of a density $\phi(\vR)$ per unit of volume in the 3-dimensional space as\footnote{In the literature one also uses an
alternative definition, which we show to be wrong in the \appr{sec/Appendix/Excess/Def}.} \cite{albano/sing}
\begin{equation}\label{eq/Excess/Definition/03}
\widehat{\phi}(\xs, \vR_{\parallel}) \equiv \frac{1}{h_{2}^{s}\,h_{3}^{s}}\,\int_{\displaystyle \xgsb}^{\displaystyle \xlsb}{dx_{1}\,h_{1}\,h_{2}\,h_{3}\,\phi^{ex}(\vR; \xs)}
\end{equation}
where
\begin{equation}\label{eq/Excess/Definition/04}
\phi^{ex}(\vR; \xs) \equiv \phi(\vR) - \phi^{g}(\vR)\,\Theta(\xs-x_{1}) - \phi^{\ell}(\vR)\,\Theta(x_{1}-\xs)
\end{equation}
Furthermore, $h_{i} \equiv h_{i}(x_{1}, \vR_{\parallel})$ are Lame coefficients for curvilinear coordinates and $h_{i}^{s} \equiv h_{i}(\xs, \vR_{\parallel})$. Given that $\phi(\vR)$ is a
density per unit of volume, excess $\widehat{\phi}(\xs, \vR_{\parallel})$ is a density per unit of surface. The excess depends on the position of the dividing surface $\xs$, which is the
coordinate of the surface in the normal direction, and the position $\vR_{\parallel}$ along the surface.


%
\subsection{Stationary state of a surface.}\label{sec/Excess/Surface}

Consider the entropy production given in \eqr{eq/Entropy/Balance/05}. All the terms but last one have the form $\vJ\spd\nabla\phi$, where according to \eqr{eq/Entropy/Balance/03a} $\nabla\spd\vJ
= 0$ and $\phi$ is some scalar function. Thus, $\vJ\spd\nabla\phi = \nabla\spd(\vJ\phi)$. We show in \appr{sec/Appendix/Excess/Gradient} that
\begin{equation}\label{eq/Excess/Surface/01}
\widehat{\vJ\spd\nabla\phi} = (J_{\perp}\phi)^{\ell} - (J_{\perp}\phi)^{g} + \widehat{\nabla_{\parallel}\spd(\vJ_{\parallel}}\phi)
\end{equation}%
where all the functions on the right hand side are evaluated at $\rs$.

For each flux in \eqr{eq/Entropy/Balance/03a} we can write $\nabla_{\perp}J_{\perp} + \nabla_{\parallel}\spd\vJ_{\parallel} = 0$. This gives an approximate relation for the order of magnitude
\begin{equation}\label{eq/Excess/Surface/04}
\frac{|\Delta_{\perp}J_{\perp}|}{\Delta x_{\perp}} \simeq \frac{|\Delta_{\parallel}J_{\parallel}|}{\Delta x_{\parallel}}
\end{equation}%
As was discussed in \cite{glav/grad1}, the interfacial region breaks the 3-dimensional isotropy of the system. In addition to a typical macroscopic size of the problem $\ell$, there exists the
microscopic size $\delta$, the surface width, which is of the order of few nanometers. There are quantities which change drastically on the distances of the order $\delta$ in the direction
perpendicular to the surface. However, the significant change of any quantity along the surface may happen only on a length scale $\ell$, which is of the order of either radii of curvature or
the system size. Because of this property of a surface, we may not expect the change of the parallel component of a flux on a macroscopic scale along the surface to be much larger then the
change of the perpendicular component of that flux on a microscopic scale through the surface. For the fluxes for which changes $|\Delta_{\perp}J_{\perp}|$ and
$|\Delta_{\parallel}J_{\parallel}|$ are of the same order of magnitude, \eqr{eq/Excess/Surface/04} takes the form $\Delta J / \delta \approx \Delta J / \ell$, which can hold only if $\Delta J =
0$, since $\delta \ll \ell$. This means that both $\Delta_{\perp}J_{\perp} = 0$ and $\Delta_{\parallel}J_{\parallel} = 0$. If $|\Delta_{\perp}J_{\perp}| \gg |\Delta_{\parallel}J_{\parallel}|$
this statement becomes even stronger.
We then may require that for a thin surface\footnote{For the special case of a system with planar surface in cartesian coordinated with all the fluxes directed perpendicular to the surface,
which is the case studied in this paper, these equations follow straightforwardly.}
\begin{subequations}\label{eq/Excess/Surface/05}
\begin{equation}\label{eq/Excess/Surface/05a}
\nabla_{\perp}J_{\perp}(\vR) = 0
\end{equation}\vspace{-2.0pc}
\begin{equation}\label{eq/Excess/Surface/05b}
\nabla_{\parallel}\spd\vJ_{\parallel}(\vR) = 0
\end{equation}
\end{subequations}
Thus, a stationary state condition $\nabla\spd\vJ=0$ has a form of \eqr{eq/Excess/Surface/05} in an interfacial region. The extrapolated fluxes $\vJ^{b}$ satisfy the same equation
\begin{subequations}\label{eq/Excess/Surface/07}
\begin{equation}\label{eq/Excess/Surface/07a}
\nabla_{\perp}J_{\perp}^{b}(\vR) = 0
\end{equation}\vspace{-2.0pc}
\begin{equation}\label{eq/Excess/Surface/07b}
\nabla_{\parallel}\spd\vJ_{\parallel}^{b}(\vR) = 0
\end{equation}
\end{subequations}
since the extrapolated flux fields also satisfy $\nabla\spd\vJ^{b} = 0$.

Both \eqr{eq/Excess/Surface/05a} and \eqr{eq/Excess/Surface/07a} are first order ordinary differential equations which depend on a constant. These constants must be the same, since according to
\eqr{eq/Excess/Definition/01} $J_{\perp}^{b}(\vR^{b,s}) = J_{\perp}(\vR^{b,s})$ at boundary points. It means that $J_{\perp}^{b}$ and $J_{\perp}$ are the same functions\footnote{Note that
\eqr{eq/Excess/Surface/02a} does not lead to the relation $J_{\perp}(\xgsb, \vR_{\parallel}) = J_{\perp}(\xlsb, \vR_{\parallel})$. \eqr{eq/Excess/Surface/02a} is the relation between values of
different functions at the same point but not the relation between values of the same function at different points. However, it follows from \eqr{eq/Excess/Surface/05a} that in curvilinear
coordinates $\partial(h_{2}h_{3}J_{\perp})/(\partial x_{\perp}) = 0$ and therefore $h_{2}h_{3}J_{\perp} = const$ but not $J_{\perp} = const$. }:
\begin{equation}\label{eq/Excess/Surface/02a}
J_{\perp}^{g}(\vR) = J_{\perp}^{\ell}(\vR) = J_{\perp}(\vR)
\end{equation}%

Consider the last term in \eqr{eq/Excess/Surface/01}. Since parallel divergences of both $\vJ_{\parallel}$ and $\vJ_{\parallel}^{b}$ are zero
\begin{equation}\label{eq/Excess/Surface/06}
\widehat{\nabla_{\parallel}\spd(\vJ_{\parallel}\phi)} = \widehat{\vJ_{\parallel}\spd\nabla_{\parallel}\phi}
\end{equation}
where we used that $\nabla_{\parallel}\Theta(x_{1}) = 0$.

Substituting \eqr{eq/Excess/Surface/02a} and \eqr{eq/Excess/Surface/06} into \eqr{eq/Excess/Surface/01} we obtain
\begin{equation}\label{eq/Excess/Surface/11}
\widehat{\vJ\spd\nabla\phi} = J_{\perp}(\phi^{\ell} - \phi^{g}) + \widehat{\vJ_{\parallel}\spd\nabla_{\parallel}\phi}
\end{equation}%
where all the functions on the right hand side are evaluated at $\rs$.

\section{Excess entropy production.}\label{sec/ExcessEntropy}

Applying \eqr{eq/Excess/Surface/11} to each term in \eqr{eq/Entropy/Balance/05} for the entropy production we obtain the general form of the excess entropy production for a surface in stationary
state
\begin{equation}\label{eq/ExcessEntropy/01}
\begin{array}{rl}
\widehat{\sigma}_{s} = & {J_{s,tot}^{\ell}}_{\perp} - {J_{s,tot}^{g}}_{\perp} + \nabla_{\parallel}\spd{\widehat{\vJ}_{s,tot}{}}_{\parallel} =
\\\\
= &\displaystyle J_{e,\,\perp}\left(\frac{1}{T^{\ell}} - \frac{1}{T^{g}}\right)%
-  \sum_{i=1}^{n}{J_{\xi_{i},\,\perp}\left(\frac{\tilde{\mu}_{i}^{\ell}}{T^{\ell}}- \frac{\tilde{\mu}_{i}^{g}}{T^{g}}\right)}  %
- \left(\frac{\pi_{\perp\beta}^{\ell}\velocity_{\beta}^{\ell}}{T^{\ell}} - \frac{\pi_{\perp\beta}^{g}\velocity_{\beta}^{g}}{T^{g}}\right) %
\\\\
&+ \displaystyle \widehat{\left(\vJ_{e,\,\parallel}\spd\nabla_{\parallel}\frac{1}{T}\right)} -
\sum_{i=1}^{n}{\widehat{\left(\vJ_{\xi_{i},\,\parallel}\spd\nabla_{\parallel}\frac{\tilde{\mu}_{i}}{T}\right)}} -
\widehat{\left(\nabla_{\parallel}\spd\frac{\pi_{\parallel\beta}\velocity_{\beta}}{T}\right)}
\end{array}
\end{equation}%
where $\tilde{\mu}_{i} \equiv \mu_{i} + \velocity^{2}/2-\vg\spd\rs$.

The next step of the analysis is to provide constitutive relations in order to relate thermodynamic forces $X_{k}$ to thermodynamic fluxes $J_{k}$ for the whole surface. This requires that the
excess entropy production has a form $\widehat{\sigma}_{s} = \sum{J_{k}X_{k}}$. However, as one can see from \eqr{eq/ExcessEntropy/01}, the terms related to fluxes along the surface do not have
this form. One has to make further assumptions on the nature of these terms to write them in this form. As our work is focused on transport into and through the surface we will not consider
non-equilibrium perturbations which are applied along the surface. This guarantees that all the terms along the surface are equal to zero. The only nonzero component of any flux $\vJ$ is
therefore $J_{\perp}$, which we will denote simply as $J$. We will furthermore restrict ourself to non-viscous fluids. The expression for the excess entropy production simplifies to the
following
\begin{equation}\label{eq/ExcessEntropy/02}
\widehat{\sigma}_{s} = J_{e}\left(\frac{1}{T^{\ell}} - \frac{1}{T^{g}}\right) -  \sum_{i=1}^{n}{J_{\xi_{i}}\left(\frac{\tilde{\mu}_{i}^{\ell}}{T^{\ell}}-
\frac{\tilde{\mu}_{i}^{g}}{T^{g}}\right)}
\end{equation}%

It is convenient to write the excess entropy production in terms of the measurable heat flux $\vJ_{q}'$, rather then the total energy flux $\vJ_{e}$, which is defined as
\begin{equation}\label{eq/ExcessEntropy/03}
\vJ_{q}' \equiv \vJ_{q} - \sum_{i=1}^{n}{h_{i}\vJ_{i}} = \vJ_{e} - \sum_{i=1}^{n}{\tilde{h}_{i}\vJ_{\xi_i}}
\end{equation}%
where we used \eqr{eq/Entropy/Balance/03} and  $\tilde{h}_{i} \equiv h_{i} + \velocity^{2}/2-\vg\spd\rs = \tilde{\mu}_{i} + Ts_{i}$, where $s_{i}$ is the partial entropy and $h_{i}$ is the
partial enthalpy. While \eqr{eq/Excess/Surface/02a} is valid for $\vJ_{e}$ and $\vJ_{\xi_i}$ it is not valid for $\vJ_{q}'$: the difference between the measurable heat fluxes extrapolated from
the gas and the liquid side is
\begin{equation}\label{eq/ExcessEntropy/04}
J_{q}^{\,\prime,\,g} - J_{q}^{\,\prime,\,\ell} = \sum_{i=1}^{n}{J_{\xi_i}(\tilde{h}_{i}^{\ell}-\tilde{h}_{i}^{g})}
\end{equation}%
In terms of measurable heat fluxes the expression for the entropy production becomes
\begin{subequations}\label{eq/ExcessEntropy/05}
\begin{equation}\label{eq/ExcessEntropy/05a}
\widehat{\sigma}_{s} = J_{q}^{\,\prime,\,g}\left(\frac{1}{T^{\ell}} - \frac{1}{T^{g}}\right) -  \sum_{i=1}^{n}{J_{\xi_{i}}\frac{1}{T^{\ell}}\left(\tilde{\mu}_{i}^{\ell}- \tilde{\mu}_{i}^{g} +
s_{i}^{g}(T^{\ell}-T^{g})\right)}
\end{equation}
\begin{equation}\label{eq/ExcessEntropy/05b}
\widehat{\sigma}_{s} = J_{q}^{\,\prime,\,\ell}\left(\frac{1}{T^{\ell}} - \frac{1}{T^{g}}\right) -  \sum_{i=1}^{n}{J_{\xi_{i}}\frac{1}{T^{g}} \left(\tilde{\mu}_{i}^{\ell}- \tilde{\mu}_{i}^{g} +
s_{i}^{\ell}(T^{\ell}-T^{g})\right)}
\end{equation}
\end{subequations}
It is important to realize that \eqr{eq/ExcessEntropy/05} are exactly equivalent to \eqr{eq/ExcessEntropy/02}. It is common to do these transformations neglecting third and higher order
contributions in the deviation from equilibrium. Such approximations were not needed here.

\eqr{eq/ExcessEntropy/05} has the form of the entropy production for the surface used in \cite{kjelstrupbedeaux/heterogeneous}. It was obtained there using the local equilibrium hypothesis,
which we have proven to be valid in \cite{glav/grad2}. In this article we have derived \eqr{eq/ExcessEntropy/05} independently, by calculating the excess of the continuous entropy production in
the gradient model.

\section{Surface transfer coefficients.}\label{sec/TransportCoef}
Consider \eqr{eq/ExcessEntropy/05} for excess entropy production which has the form
\begin{equation}\label{eq/TransportCoef/001}
\widehat{\sigma}_{s} = J_{q}^{\,\prime}X_{q} -  \sum_{i=1}^{n}{J_{\xi_{i}}X_{i}}
\end{equation}
We will use the form \eqref{eq/TransportCoef/001} further, specifying the explicit expressions for fluxes and forces where needed. Following the common procedure we write the linear force-flux
relations for a given entropy production:
\begin{equation}\label{eq/TransportCoef/002}
\begin{array}{rl}
X_{q} &= R_{qq}(T_{eq},\psi_{eq})\,J_{q}^{\,\prime} + \sum_{i=1}^{n}{R_{qi}(T_{eq},\psi_{eq})\,J_{\xi_{i}}} \\\\
-X_{j} &= R_{jq}(T_{eq},\psi_{eq})\,J_{q}^{\,\prime} + \sum_{i=1}^{n}{R_{ji}(T_{eq},\psi_{eq})\,J_{\xi_{i}}}
\end{array}
\end{equation}
As these relations are true only to the linear order in perturbations, we must keep only linear contributions in all terms. It means that all the resistivities in \eqr{eq/TransportCoef/002} are
functions of only equilibrium temperature $T_{eq}$ and equilibrium chemical potential difference $\psi_{eq}$ around which a particular perturbation is performed. They do not depend on the nature
of the perturbation.

Consider the following matrix notation of the above quantities
\begin{equation}\label{eq/TransportCoef/10}
\mathrm{X} \equiv \begin{pmatrix}[c] X_{q} \\ -X_{1}\\ \vdots \\ -X_{n} \end{pmatrix}, \quad%
\mathrm{R} \equiv \begin{pmatrix} R_{qq} & R_{q1} & \ldots & R_{qn} \\ R_{1q} & R_{11} & \ldots & R_{1n} \\ \vdots & \vdots & \ddots & \vdots \\ R_{nq} & R_{n1} & \ldots & R_{nn} \end{pmatrix}, \quad%
\mathrm{J} \equiv \begin{pmatrix}[c] J_{q}^{\,\prime} \\ J_{\xi_1} \\ \vdots\\ J_{\xi_n} \end{pmatrix} %
\end{equation}
Let $\ss$ indicate a measure of a non-equilibrium perturbation, so that $\mathrm{X} = \mathrm{X}(\ss)$ and $\mathrm{J} = \mathrm{J}(\ss)$. Then \eqr{eq/TransportCoef/002} can be written in a
matrix form as
\begin{equation}\label{eq/TransportCoef/11}
\mathrm{X}(\ss) = \mathrm{R}(T_{eq}, \psi_{eq})\spd\mathrm{J}(\ss)
\end{equation}
For big values of $\ss$ \eqr{eq/TransportCoef/11} is not correct, since big perturbations are not described by the linear theory. As we decrease $\ss$, the accuracy of \eqr{eq/TransportCoef/11}
increases and in the limit $\ss \rightarrow 0$ becomes exact. It means that \eqr{eq/TransportCoef/11} should be understood as
\begin{equation}\label{eq/TransportCoef/12}
\lim_{\ss \rightarrow 0}\mathrm{X}(\ss) = \mathrm{R}(T_{eq}, \psi_{eq}) \spd \lim_{\ss \rightarrow 0}\mathrm{J}(\ss) \\
\end{equation}

One should not write \eqr{eq/TransportCoef/12} in the form $\mathrm{X}(0) = \mathrm{R}(T_{eq}, \psi_{eq})\spd\mathrm{J}(0)$ however, as both $\mathrm{X}(0)$ and $\mathrm{J}(0)$ contain only
zeroes and such an expression makes no sense. Even though $\mathrm{X}(\ss)$ and $\mathrm{J}(\ss)$ are continuous functions of $\ss$, one should write $\lim_{\ss \rightarrow 0}\mathrm{X}(\ss)$
and $\lim_{\ss \rightarrow 0}\mathrm{J}(\ss)$ instead of $\mathrm{X}(0)$ and $\mathrm{J}(0)$ respectively. In practice there exists a particular measure $\ss_{eq}$ of a perturbation, such that
for all $\ss < \ss_{eq}$ \eqr{eq/TransportCoef/11} is satisfied with a satisfactory accuracy.

One should also note that the accuracy of a particular numerical procedure may limit the validity of \eqr{eq/TransportCoef/12} as well. All the non-equilibrium profiles and therefore forces and
fluxes are calculated by solving the system of differential equations numerically with some particular accuracy. If a perturbation rate is lower then this accuracy, say $\ss_{num}$, the data
obtained from the numerical procedure are not reliable. Performing a numerical analysis we must therefore replace the limiting value $0$ by $\ss_{num}$ in \eqr{eq/TransportCoef/12}.

As shown by Onsager \cite{onsager/rec}, the cross coefficients must be the same. We therefore have
\begin{equation}\label{eq/TransportCoef/08}
\begin{array}{rl}
R_{qi} &= R_{iq} \\
R_{ji} &= R_{ij} \\
\end{array}
\end{equation}
\subsection{Gas- and liquid- side transport coefficients.}\label{sec/TransportCoef/GL}

For each of \eqr{eq/ExcessEntropy/05} one might associate the forces with
\begin{equation}\label{eq/TransportCoef/00}
\begin{array}{rl}
X_{q}^{g} = X_{q}^{\ell} &\equiv \displaystyle \frac{1}{T^{\ell}} - \frac{1}{T^{g}} \\\\
X_{j}^{g} & \equiv \displaystyle \frac{1}{T^{\ell}}\left(\tilde{\mu}_{j}^{\ell}- \tilde{\mu}_{j}^{g} + s_{j}^{g}(T^{\ell}-T^{g})\right) \\\\
X_{j}^{\ell} &\equiv \displaystyle \frac{1}{T^{g}}\left(\tilde{\mu}_{j}^{\ell}- \tilde{\mu}_{j}^{g} + s_{j}^{\ell}(T^{\ell}-T^{g})\right)
\end{array}
\end{equation}
As we are in the context of the linear theory, however, we must linearize these forces with respect to the perturbation and discard all higher order terms. Leaving them would not increase
accuracy but may affect the consistency of the linear theory. We therefore get the following phenomenological relations up to the linear order
\begin{equation}\label{eq/TransportCoef/01}
\begin{array}{rl}
\displaystyle \frac{1}{T^{\ell}} - \frac{1}{T^{g}} &= R_{qq}^{g}\,J_{q}^{\,\prime,\,g} + \sum_{i=1}^{n}{R_{qi}^{g}\,J_{\xi_{i}}} \\\\
-\displaystyle \frac{1}{T_{eq}}\left(\tilde{\mu}_{j}^{\ell}- \tilde{\mu}_{j}^{g} + s_{j,\,eq}^{g}(T^{\ell}-T^{g})\right) &= R_{jq}^{g}\,J_{q}^{\,\prime,\,g} +
\sum_{i=1}^{n}{R_{ji}^{g}\,J_{\xi_{i}}}
\end{array}
\end{equation}
and
\begin{equation}\label{eq/TransportCoef/02}
\begin{array}{rl}
\displaystyle \frac{1}{T^{\ell}} - \frac{1}{T^{g}} &= R_{qq}^{\ell}\,J_{q}^{\,\prime,\,\ell} + \sum_{i=1}^{n}{R_{qi}^{\ell}\,J_{\xi_{i}}} \\\\
-\displaystyle \frac{1}{T_{eq}}\left(\tilde{\mu}_{j}^{\ell}- \tilde{\mu}_{j}^{g} + s_{j,\,eq}^{\ell}(T^{\ell}-T^{g})\right) &= R_{jq}^{\ell}\,J_{q}^{\,\prime,\,\ell} +
\sum_{i=1}^{n}{R_{ji}^{\ell}\,J_{\xi_{i}}}
\end{array}
\end{equation}
Here and after we omit arguments $(T_{eq},\psi_{eq})$ as long as it does not lead to confusion.

The measurable heat fluxes are related by \eqr{eq/ExcessEntropy/04} which after linearization takes the following form
\begin{equation}\label{eq/TransportCoef/03}
J_{q}^{\,\prime,\,g} - J_{q}^{\,\prime,\,\ell} = \sum_{i=1}^{n}{J_{\xi_i}\left(\tilde{h}_{i,\,eq}^{\ell}-\tilde{h}_{i,\,eq}^{g}\right)}
\end{equation}%
Comparing \eqr{eq/TransportCoef/01} and \eqr{eq/TransportCoef/02} and using \eqr{eq/TransportCoef/03} we get the following relations between the coefficients associated with the gas an liquid
measurable heat fluxes to linear order
\begin{equation}\label{eq/TransportCoef/04}
\begin{array}{rl}
R_{qq}^{\ell} &= R_{qq}^{g}\\\\
R_{qi}^{\ell} - {h}_{i,\,eq}^{\ell}\,R_{qq}^{\ell} &= R_{qi}^{g} - {h}_{i,\,eq}^{g}\,R_{qq}^{g} \\\\
R_{iq}^{\ell} - {h}_{i,\,eq}^{\ell}\,R_{qq}^{\ell} &= R_{iq}^{g} - {h}_{i,\,eq}^{g}\,R_{qq}^{g} \\\\
R_{ji}^{\ell} - {h}_{i,\,eq}^{\ell}\,R_{jq}^{\ell} - {h}_{j,\,eq}^{\ell}\,R_{qi}^{\ell} + {h}_{i,\,eq}^{\ell}\,{h}_{j,\,eq}^{\ell}\,R_{qq}^{\ell} &= %
R_{ji}^{g} - {h}_{i,\,eq}^{g}\,R_{jq}^{g} - {h}_{j,\,eq}^{g}\,R_{qi}^{g} + {h}_{i,\,eq}^{g}\,{h}_{j,\,eq}^{g}\,R_{qq}^{g}\\\\
\end{array}
\end{equation}
where we took into account that $\tilde{\mu}_{i,\,eq}^{g} = \tilde{\mu}_{i,\,eq}^{\ell}$ and $\tilde{h}_{i,\,eq}^{g}-\tilde{h}_{i,\,eq}^{\ell} = h_{i,\,eq}^{g}-h_{i,\,eq}^{\ell}$.

The coefficients on the one side determine uniquely the coefficients on the other side, having the values of jumps across the surface of the extrapolated enthalpies. It follows from
\eqr{eq/TransportCoef/04} that
\begin{equation}\label{eq/TransportCoef/04a}
\begin{array}{rl}
R_{qq}^{\ell} &= R_{qq}^{g}\\\\
R_{qi}^{\ell} &= R_{qi}^{g} - ({h}_{i,\,eq}^{g} - {h}_{i,\,eq}^{\ell})\,R_{qq}^{g} \\\\
R_{iq}^{\ell} &= R_{iq}^{g} - ({h}_{i,\,eq}^{g} - {h}_{i,\,eq}^{\ell})\,R_{qq}^{g} \\\\
R_{ji}^{\ell} &= R_{ji}^{g} - ({h}_{i,\,eq}^{g} - {h}_{i,\,eq}^{\ell})\,R_{jq}^{g} - ({h}_{j,\,eq}^{g} - {h}_{j,\,eq}^{\ell})\,R_{iq}^{g} + ({h}_{i,\,eq}^{g} - {h}_{i,\,eq}^{\ell})\,({h}_{j,\,eq}^{g} - {h}_{j,\,eq}^{\ell})\,R_{qq}^{g}\\\\
\end{array}
\end{equation}
One can notice, that symmetry of $R^{g}$ coefficients implies the same symmetry of $R^{\ell}$ coefficients and vice versa.

To avoid confusion we recall that for the temperature $T$, chemical potential $\mu_{j}$ and the partial entropy $s_{j}$ the superscript $g$ or $\ell$ means the value of the corresponding
function extrapolated from either gas or liquid to the interfacial region and evaluated at a particular dividing surface $\xs$. We do not indicate which dividing surface is used as it is
irrelevant for the present analysis. In contrast, $J_{q}^{\,\prime\,g} \equiv J_{e} - \sum_{i=1}^{n}{\tilde{h}_{i}(\xgsb)\,J_{\xi_i}}$ where the partial enthalpy is evaluated the the gas-surface
boundary $\xgsb$. To the linear order however $\tilde{h}_{i}(\xgsb) = \tilde{h}_{i,\,eq}(\xgsb) = \tilde{h}_{i,\,eq}^{g}(\xs) \equiv \tilde{h}_{i,\,eq}^{g}$ and since $J_{e}$ and $J_{\xi_i}$ are
constants, $J_{q}^{\,\prime\,g}$ may be considered as the flux, evaluated at a dividing surface $\xs$. Furthermore, superscript $g$ or $\ell$ for the resistivity $R^{g}$ or $R^{\ell}$ neither
indicate any position nor the extrapolated resistivity coefficient. It indicates the measurable heat flux with which the given resistivity coefficient is associated, either
$J^{\,\prime,\,g}_{q}$ or $J^{\,\prime,\,\ell}_{q}$ respectively.

\subsection{Mass and molar transport coefficients.}\label{sec/TransportCoef/MM}

In applications it is common to use the mass flux of the components and the partial molar thermodynamic quantities, like, for instance, partial molar entropy. The above equations should use
either molar fluxes and partial molar thermodynamic quantities or mass fluxes and partial mass thermodynamic quantities. The transport coefficients are different for different choices. Consider
\eqr{eq/TransportCoef/001} for excess entropy production. The thermodynamic forces $X_{i}$ depend on partial thermodynamic quantities. We introduce therefore $X_{i}^{m}$ as a force which uses
partial mass quantities and $X_{i}^{\nu}$ as a force which uses partial molar quantity. Furthermore, let $J_{\xi_{i}} \equiv \rho_{i}\velocity_{i}$ and $J_{\zeta_{i}} \equiv c_{i}\velocity_{i}$
be the mass and molar flux respectively of the $i$-th component. As $X_{i}^{m} = X_{i}^{\nu} / M_{i}$ and $J_{\xi_{i}} = J_{\zeta_{i}} M_{i}$, where $M_{i}$ is the molar mass of the $i$-th
component, the excess entropy production becomes
\begin{equation}\label{eq/TransportCoef/05}
\widehat{\sigma}_{s} = J_{q}^{\,\prime}X_{q} -  \sum_{i=1}^{n}{J_{\xi_{i}}X_{i}^{m}} = J_{q}^{\,\prime}X_{q} -  \sum_{i=1}^{n}{J_{\zeta_{i}}X_{i}^{\nu}}
\end{equation}
The force-flux relations become
\begin{subequations}\label{eq/TransportCoef/06}
\begin{equation}\label{eq/TransportCoef/06a}
\begin{array}{rl}
\vspace{-0.5pc}%
X_{q} &= R_{qq}^{m}\,J_{q}^{\,\prime} + \sum_{i=1}^{n}{R_{qi}^{m}\,J_{\xi_{i}}} \\\\
-X_{j}^{m} &= R_{jq}^{m}\,J_{q}^{\,\prime} + \sum_{i=1}^{n}{R_{ji}^{m}\,J_{\xi_{i}}}
\end{array}
\end{equation}
and
\begin{equation}\label{eq/TransportCoef/06b}
\begin{array}{rl}
\vspace{-0.5pc}%
X_{q} &= R_{qq}^{\nu}\,J_{q}^{\,\prime} + \sum_{i=1}^{n}{R_{qi}^{\nu}\,J_{\zeta_{i}}} \\\\
-X_{j}^{\nu} &= R_{jq}^{\nu}\,J_{q}^{\,\prime} + \sum_{i=1}^{n}{R_{ji}^{\nu}\,J_{\zeta_{i}}}
\end{array}
\end{equation}
\end{subequations}
where the corresponding superscript for the resistivity coefficient indicates the association with the mass or molar quantities. These transport coefficients are related in the following way
\begin{equation}\label{eq/TransportCoef/07}
\begin{array}{rl}
\vspace{-0.5pc}%
R_{qq}^{\nu} &= R_{qq}^{m}\\\\
\vspace{-0.5pc}%
R_{qi}^{\nu} &= M_{i}\,R_{qi}^{m}\\\\
\vspace{-0.5pc}%
R_{iq}^{\nu} &= M_{i}\,R_{iq}^{m}\\\\
R_{ji}^{\nu} &= M_{j}M_{i}\,R_{ji}^{m}\\
\end{array}
\end{equation}
\section{Two component mixture.}\label{sec/TwoComponent}

We now restrict ourselves to the mixture considered in \cite{glav/grad2}. We consider the mixture of cyclohexane (1st component) and $n$-hexane (2nd component) in a box with gravity directed
along axes $x$ from left to right. The gas phase is therefore in the left part of the box and the liquid is in the right part of the box. The surface is planar.

We have the following expression for the excess entropy production
\begin{equation}\label{eq/TwoComponent/00}
\widehat{\sigma}_{s} = J_{q}^{\,\prime}X_{q} -  J_{\xi_{1}}X_{1} - J_{\xi_{2}}X_{2}
\end{equation}
and force-flux relations \eqref{eq/TransportCoef/11} with
\begin{equation}\label{eq/TwoComponent/03}
\mathrm{X} \equiv \begin{pmatrix}[c] X_{q} \\ -X_{1} \\ -X_{2} \end{pmatrix}, \quad%
\mathrm{R} \equiv \begin{pmatrix} R_{qq} & R_{q1} & R_{q2} \\ R_{1q} & R_{11} & R_{12} \\ R_{2q} & R_{21} & R_{22} \end{pmatrix}, \quad%
\mathrm{J} \equiv \begin{pmatrix}[c] J_{q}^{\,\prime} \\ J_{\xi_1} \\ J_{\xi_2} \end{pmatrix} %
\end{equation}

We will also use an alternative expression for the excess entropy production, which uses the total mass flux $J_{m} = J_{\xi_{1}} + J_{\xi_{2}}$ and the flux of one of the components, say
$J_{\xi_{1}} \equiv J_{\xi}$:
\begin{equation}\label{eq/TwoComponent/10}
\widehat{\sigma}_{s} = J_{q}^{\,\prime}X_{q} -  J_{\xi}X_{\xi} - J_{m}X_{m}
\end{equation}
where $X_{\xi} \equiv X_{1}-X_{2}$ and $X_{m} \equiv X_{2}$. The resulting force-flux relations \eqref{eq/TransportCoef/11} have the following terms
\begin{equation}\label{eq/TwoComponent/03a}
\mathrm{X} \equiv \begin{pmatrix}[c] X_{q} \\ -X_{\xi} \\ -X_{m} \end{pmatrix}, \quad%
\mathrm{R} \equiv \begin{pmatrix} R_{qq} & R_{q\xi} & R_{qm} \\ R_{\xi q} & R_{\xi\xi} & R_{\xi m} \\ R_{mq} & R_{m\xi} & R_{mm} \end{pmatrix}, \quad%
\mathrm{J} \equiv \begin{pmatrix}[c] J_{q}^{\,\prime} \\ J_{\xi} \\ J_{m} \end{pmatrix} %
\end{equation}
where the coefficients from \eqr{eq/TwoComponent/03} are related to the coefficients from \eqr{eq/TwoComponent/03a} as
\begin{equation}\label{eq/TwoComponent/12}
\begin{array}{ll}
R_{q1} = R_{q\xi} + R_{qm}  &\quad\quad R_{11} = -R_{m\xi} - R_{\xi m} + R_{mm} + R_{\xi\xi} \\
R_{1q} = R_{\xi q} + R_{mq} &\quad\quad R_{22} = R_{mm} \\
R_{q2} = R_{qm}             &\quad\quad R_{12} = R_{mm} - R_{\xi m} \\
R_{2q} = R_{mq}             &\quad\quad R_{21} = R_{mm} - R_{m\xi}\\
\end{array}
\end{equation}

Having the numerical solution for a particular non-equilibrium stationary state we know all the fluxes $\mathrm{J}$ and forces $\mathrm{X}$ used in \eqr{eq/TransportCoef/11}: the constant fluxes
are obtained directly from the non-equilibrium solution and the extrapolated bulk profiles are obtained using the procedure described in \cite{glav/grad2}. On the other hand we know only the
local resistivities but not the resistivities $\mathrm{R}$ of the whole surface.

We now consider the inverse problem: to determine the transport coefficients for the whole surface having the non-equilibrium solution. As one can see, \eqr{eq/TransportCoef/11} has 9 unknown
resistivities\footnote{Solving the inverse problem we have to proof the Onsager reciprocal relations rather then impose them.} and only 3 equations. It is therefore not possible to determine all
the transport coefficients uniquely having only one stationary state solution. In order to incorporate more equations we need to consider other non-equilibrium stationary solutions which are
independent of the previous. As the transport coefficients depend only on equilibrium unperturbed state but not on non-equilibrium perturbations, considering different perturbations around the
same equilibrium state we will get missing data. We must ensure however, that a given perturbation is small enough to be described by linear-order equations. This would require for instance
\eqr{eq/TransportCoef/08} and \eqr{eq/TransportCoef/04} to be true. There are more constraints to be fulfilled which will be discussed in \secr{sec/Results}.

The non-equilibrium solution uses following profiles for local resistivities (see \cite{glav/grad2} for details)
\begin{equation}\label{eq/Equations/Phenomenological/03}
\begin{array}{rl}
r_{qq}(x) &= r_{qq}^{g} + (r_{qq}^{\ell}-r_{qq}^{g})\,q_{0}(x) + \alpha_{qq}(r_{qq}^{\ell}+r_{qq}^{g})\,q_{1}(x)\\
\\
r_{q1}(x) &= r_{q1}^{g} + (r_{q1}^{\ell}-r_{q1}^{g})\,q_{0}(x) + \alpha_{q1}(r_{q1}^{\ell}+r_{q1}^{g})\,q_{1}(x)\\
\\
r_{11}(x) &= r_{11}^{g} + (r_{11}^{\ell}-r_{11}^{g})\,q_{0}(x) + \alpha_{11}(r_{11}^{\ell}+r_{11}^{g})\,q_{1}(x)\\
\end{array}
\end{equation}
where $q_{0}(x)$ and $q_{1}(x)$ are modulatory curves for resistivity profiles which depend only on density profiles and their first derivatives. For each resistivity profile $r^{g}$ and
$r^{\ell}$ are the equilibrium coexistence resistivities of the gas and liquid phase respectively. Coefficients $\alpha_{qq}$, $\alpha_{q1}$, $\alpha_{11}$ control the size of the peak in the
resistivity profiles in the interfacial region. The non-equilibrium stationary state depends on the values of these coefficients. The surface resistivity coefficients $\mathrm{R}$ will therefore
depend on these coefficients as on parameters, $\mathrm{R} = \mathrm{R}(\alpha_{qq}, \alpha_{q1}, \alpha_{11})$, which we will investigate.

\section{Methods to obtain resistivities.}\label{sec/Methods}

We determine the transport coefficients from three different methods: from a "perturbation cell" method\footnote{This method was first used by Johannessen et. al. in \cite{bedeaux/vdW/III} for
one-component system. Here we discuss the grounds for the legitimacy of this procedure and generalize it to mixtures.}, from an experimental-like procedure and from kinetic theory.

\subsection{Perturbation cell.}\label{sec/TwoComponent/Equilibrium}

Consider a stationary state which is perturbed from equilibrium by setting the temperature of the liquid\footnote{One should not confuse $T(\xl)$ with $T^{\ell}$. The former is the actual
temperature at $x=\xl$, i.e. at the box boundary on the liquid side. The latter is the temperature extrapolated from the liquid phase to the interfacial region and taken at $x=\xs$, i.e. at the
dividing surface.} $T(\xl) = (1+\beta_{T})T_{eq}$, the pressure of the gas $p(\xg) = (1+\beta_{p})p_{eq}$ and the mole fraction of the liquid $\zeta^{\ell}(\xl) =
(1+\beta_{\zeta})\zeta^{\ell}_{eq}$ independently. The resulting non-equilibrium state is therefore a function of parameters $\beta$:
\begin{equation}\label{eq/TwoComponent/04}
\mathrm{X}(\beta_{T}, \beta_{p}, \beta_{\zeta}) = \mathrm{R}(T_{eq}, \psi_{eq})\spd\mathrm{J}(\beta_{T}, \beta_{p}, \beta_{\zeta})
\end{equation}
where $\mathrm{X}$, $\mathrm{J}$ and $\mathrm{R}$ are given by \eqr{eq/TwoComponent/03}. Consider the following set of 8 independent non-equilibrium perturbations:
\begin{equation}\label{eq/TwoComponent/07}
\begin{array}{rrrrcrcrrrr}
\mathrm{X}(&\beta,& \beta,& \beta) &=& \mathrm{R}(T_{eq}, \psi_{eq})&\spd&\mathrm{J}(&\beta,& \beta,& \beta) \\
\mathrm{X}(&\beta,& -\beta,& \beta) &=& \mathrm{R}(T_{eq}, \psi_{eq})&\spd&\mathrm{J}(&\beta,& -\beta,& \beta) \\
\mathrm{X}(&-\beta,& \beta,& \beta) &=& \mathrm{R}(T_{eq}, \psi_{eq})&\spd&\mathrm{J}(&-\beta,& \beta,& \beta) \\
\mathrm{X}(&-\beta,& -\beta,& \beta) &=& \mathrm{R}(T_{eq}, \psi_{eq})&\spd&\mathrm{J}(&-\beta,& -\beta,& \beta) \\
\mathrm{X}(&\beta,& \beta,& -\beta) &=& \mathrm{R}(T_{eq}, \psi_{eq})&\spd&\mathrm{J}(&\beta,& \beta,& -\beta) \\
\mathrm{X}(&\beta,& -\beta,& -\beta) &=& \mathrm{R}(T_{eq}, \psi_{eq})&\spd&\mathrm{J}(&\beta,& -\beta,& -\beta) \\
\mathrm{X}(&-\beta,& \beta,& -\beta) &=& \mathrm{R}(T_{eq}, \psi_{eq})&\spd&\mathrm{J}(&-\beta,& \beta,& -\beta) \\
\mathrm{X}(&-\beta,& -\beta,& -\beta) &=& \mathrm{R}(T_{eq}, \psi_{eq})&\spd&\mathrm{J}(&-\beta,& -\beta,& -\beta) \\
\end{array}
\end{equation}
Consider now the $3 \times 8$ matrices $\mathfrak{X}$ and $\mathfrak{J}$ which contain 8 column vectors $\mathrm{X}$ and $\mathrm{J}$ respectively for each non-equilibrium perturbation specified
above. For these perturbations $\mathfrak{X} = \mathfrak{X}(\beta)$ and $\mathfrak{J} = \mathfrak{J}(\beta)$ are the functions only on one parameter $\beta$. It follows from
\eqr{eq/TwoComponent/07} that
\begin{equation}\label{eq/TwoComponent/08}
\mathfrak{X}(\beta) = \mathrm{R}(T_{eq}, \psi_{eq}) \spd \mathfrak{J}(\beta)
\end{equation}
As it was discussed in \secr{sec/TransportCoef}, for practical purposes these limits should be calculated as the values of corresponding matrices at very small but finite value of $\beta$.
$\mathrm{R}(T_{eq}, \psi_{eq})$ depends therefore on $\beta$ and we will keep it as an argument. From \eqr{eq/TwoComponent/08} we obtain
\begin{equation}\label{eq/TwoComponent/09}
\mathrm{R}(T_{eq}, \psi_{eq}; \beta) = \left(\mathfrak{X}(\beta)\spd\mathfrak{J}^{T}(\beta)\right)\spd \left(\mathfrak{J}(\beta)\spd\mathfrak{J}^{T}(\beta)\right)^{-1} \\
\end{equation}
where superscript ${}^{T}$ means the matrix transpose and ${}^{-1}$ means the inverted matrix.

We note, that in order to obtain the resistivity matrix $\mathrm{R}$ uniquely, it is sufficient in principle to impose any 3 non-equilibrium perturbations which have sufficiently small
perturbation parameters $\beta_{T}$, $\beta_{p}$ and $\beta_{\zeta}$. As \textit{each} of $\beta_{T}$, $\beta_{p}$ and $\beta_{\zeta}$ goes to zero the resistivity matrix will go to
$\mathrm{R}(T_{eq},\psi _{eq})$ as fast as \textit{all} $\beta_{T}$, $\beta_{p}$ and $\beta_{\zeta}$ go to zero. The method presented above makes the resistivity matrix converge to
$\mathrm{R}(T_{eq},\psi _{eq})$ as fast as $\beta^{2}$ goes to zero, however. This is achieved by using 8 perturbations at the "corners" of a three-dimensional "perturbation cell", so changing
$\beta $ to $-\beta $ would not change the "perturbation cell" and the resulting $\mathrm{R}$.

Because of using 8 perturbations instead of 3, there are 5 superfluous perturbations which make the system of equations \eqref{eq/TwoComponent/08} to be overdetermined. Contracting both sides of
\eqr{eq/TwoComponent/08} with $\mathfrak{J}^{T}$ we actually average all the perturbations which are spread around $T_{eq}$ and $\psi_{eq}$ in the least square sense. As the components of
$\mathfrak{J}$ matrix are linearly independent, this guaranteers the matrix $\mathfrak{J}\spd\mathfrak{J}^{T}$ to be invertible. Thus, the inverse matrix
$(\mathfrak{J}\spd\mathfrak{J}^{T})^{-1}$ exists and \eqr{eq/TwoComponent/09} is mathematically legitimate. In the numerical procedure the expression on the right hand side of
\eqr{eq/TwoComponent/09} is obtained using Matlab matrix division $\mathfrak{X}/\mathfrak{J}$.

\subsection{Experiment-like procedure.}\label{sec/TwoComponent/Experiment}

In experiments it is convenient to measure the corresponding coefficients by keeping zero mass fluxes through the system. It is also convenient to work with the total mass flux $J_{m}$ and the
flux of one of the components $J_{\xi}$, rather then with fluxes of each component separately\footnote{One of the reasons for this is that it is hard to make only $J_{\xi_{1}} = 0$, keeping
$J_{\xi_{2}}$ finite.}, $J_{\xi_{1}}$ and $J_{\xi_{2}}$.

Consider a stationary state which is perturbed from equilibrium by setting the temperature of the liquid $T(\xl) = (1+\beta_{T})T_{eq}$. The second perturbation condition is either $J_{\xi} = 0$
or $\zeta^{\ell}(\xl) = \zeta^{\ell}_{eq}$ and we introduce the perturbation parameter $\beta_{\xi}$ which is 0 in the former case and 1 in the latter one. The third perturbation condition is
either $J_{m} = 0$ or $p(\xg) = p_{eq}$ and the corresponding perturbation parameter $\beta_{m}$ is 0 or 1 respectively. The resulting non-equilibrium state is therefore a function of 3
parameters:
\begin{equation}\label{eq/TwoComponent/13}
\mathrm{X}(\beta_{T}, \beta_{\xi}, \beta_{m}) = \mathrm{R}(T_{eq}, \psi_{eq})\spd\mathrm{J}(\beta_{T}, \beta_{\xi}, \beta_{m})
\end{equation}
where $\mathrm{X}$, $\mathrm{J}$ and $\mathrm{R}$ are given by \eqr{eq/TwoComponent/03a}. Consider the following set of 3 independent non-equilibrium perturbations:
\begin{equation}\label{eq/TwoComponent/15}
\begin{array}{rl}
\mathrm{X}(\beta, 0, 0) &= \mathrm{R}(T_{eq}, \psi_{eq})\spd\mathrm{J}(\beta, 0, 0) \\
\mathrm{X}(\beta, 1, 0) &= \mathrm{R}(T_{eq}, \psi_{eq})\spd\mathrm{J}(\beta, 1, 0) \\
\mathrm{X}(\beta, 1, 1) &= \mathrm{R}(T_{eq}, \psi_{eq})\spd\mathrm{J}(\beta, 1, 1) \\
\end{array}
\end{equation}
\begin{subequations}\label{eq/TwoComponent/16}
From the first of \eqr{eq/TwoComponent/15} we find
\begin{equation}\label{eq/TwoComponent/16a}
\begin{array}{rcrl}
R_{qq\;}    (T_{eq}, \psi_{eq}) &=& X_{q,\;\,00}    &/\, J_{q,\,00}^{\,\prime} \\
R_{\xi q\;} (T_{eq}, \psi_{eq}) &=& X_{\xi,\;\,00}  &/\, J_{q,\,00}^{\,\prime}\\
R_{mq}      (T_{eq}, \psi_{eq}) &=& X_{m,\,00}      &/\, J_{q,\,00}^{\,\prime}\\
\end{array}
\end{equation}
where we use subscripts ${}_{\beta_{\xi}\beta_{m}}$ instead the functional dependence $(\beta, \beta_{\xi}, \beta_{m})$ for simplicity of notation. From the second of \eqr{eq/TwoComponent/15} we
find
\begin{equation}\label{eq/TransportCoef/16b}
\begin{array}{rcrl}
R_{q\xi\;}  (T_{eq}, \psi_{eq}) &=& \left(X_{q,\;\,10}     - R_{qq\;}   (T_{eq}, \psi_{eq})\,J_{q,\,10}^{\,\prime}\right)    &/\, J_{\xi,\,10} \\
R_{\xi\xi\;}(T_{eq}, \psi_{eq}) &=& \left(X_{\xi,\;\,10}   - R_{\xi q\;}(T_{eq}, \psi_{eq})\,J_{q,\,10}^{\,\prime}\right) &/\, J_{\xi,\,10} \\
R_{m\xi}    (T_{eq}, \psi_{eq}) &=& \left(X_{m,\,10}       - R_{mq}     (T_{eq}, \psi_{eq})\,J_{q,\,10}^{\,\prime}\right)    &/\, J_{\xi,\,10} \\
\end{array}
\end{equation}
The values $\mathrm{X}_{10}$ and $\mathrm{J}_{10}$ are found directly from the calculations and the values of $R_{qq}(T_{eq}, \psi_{eq})$, $R_{\xi q}(T_{eq}, \psi_{eq})$ and $R_{mq}(T_{eq},
\psi_{eq})$ are those which are found in \eqr{eq/TwoComponent/16a}, given the perturbation rate $\beta$ is small enough. From the third of \eqr{eq/TwoComponent/15} we find
\begin{equation}\label{eq/TransportCoef/16c}
\begin{array}{rcrl}
R_{qm\;}    (T_{eq}, \psi_{eq}) &=& \left(X_{q,\;\,11}\;   - R_{qq\;}     (T_{eq}, \psi_{eq})\,J_{q,\,11}^{\,\prime} - R_{q\xi\;}  (T_{eq}, \psi_{eq})\,J_{\xi,\,11} \right) &/\, J_{m,\,11} \\
R_{\xi m\;} (T_{eq}, \psi_{eq}) &=& \left(X_{\xi,\;\,11}\; - R_{\xi q\;}  (T_{eq}, \psi_{eq})\,J_{q,\,11}^{\,\prime} - R_{\xi\xi\;}(T_{eq}, \psi_{eq})\,J_{\xi,\,11} \right) &/\, J_{m,\,11} \\
R_{mm}      (T_{eq}, \psi_{eq}) &=& \left(X_{m,\,11}       - R_{mq}       (T_{eq}, \psi_{eq})\,J_{q,\,11}^{\,\prime} - R_{m\xi}    (T_{eq}, \psi_{eq})\,J_{\xi,\,11} \right) &/\, J_{m,\,11} \\
\end{array}
\end{equation}
Again, all the quantities on the right hand side of \eqr{eq/TransportCoef/16c} are known and we therefore can find the remaining resistivities.
\end{subequations}
\subsection{Comparison to kinetic theory.}\label{sec/KineticTheory}

According to \cite[p. 180]{kjelstrupbedeaux/heterogeneous} kinetic theory gives the following expressions for the surface transport coefficients for two component mixture
\begin{equation}\label{eq/KineticTheory/01}
\begin{array}{rl}
R_{qq}^{g,\,\nu}(T, \psi) &= 4\,R_{O}(T, \psi)\left\{1 + \displaystyle \frac{104}{25\pi}\left(\frac{w_{1}^{2}}{\varsigma_{1}} + \frac{w_{2}^{2}}{\varsigma_{2}}\right)\right\} \\\\
R_{qi}^{g,\,\nu}(T, \psi) = R_{iq}^{g,\,\nu}(T, \psi) &= 2RT\,R_{O}(T, \psi)\left\{1 + \displaystyle \frac{16}{5\pi}\,\frac{w_{i}}{\varsigma_{i}}\right\} \\\\
R_{ij}^{g,\,\nu}(T, \psi) &= (RT)^{2}\,R_{O}(T, \psi)\left\{1 + 32\,\delta_{ij}\,\displaystyle\frac{1}{\varsigma_{i}}\left(\frac{1}{\sigma_{i}} + \frac{1}{\pi} - \frac{3}{4}\right)\right\}
\end{array}
\end{equation}
where
\begin{equation}\label{eq/KineticTheory/01}
\begin{array}{rl}
R_{O}(T, \psi) &\equiv \displaystyle 2^{-9/2}\,\sqrt{\pi}\,R\,(RT)^{-5/2} \big(c^{g}_{1}/\sqrt{M_{1}} + c^{g}_{2}/\sqrt{M_{2}} \big)^{-1}\\
\\
\varsigma_{i}(T, \psi) &\equiv \displaystyle \big(c^{g}_{i}/\sqrt[4]{M_{i}}\big)/\big(c^{g}_{1}/\sqrt[4]{M_{1}} + c^{g}_{2}/\sqrt[4]{M_2}\big)\\
\\
w_{i}(T, \psi) &\equiv \displaystyle \lambda_{i}/(\lambda_{1}+\lambda_{2})
\end{array}
\end{equation}
where $R$ is universal gas constant, $\lambda_{i}$ and $c^{g}_{i,eq}$ are the thermal conductivity and the gas coexistence concentration of $i$-th component respectively. $\sigma_{i}$ is the
condensation coefficient, which is parameter in this theory and $\delta_{ij}$ is the Kroneker symbol.

\section{Results.}\label{sec/Results}

Using the procedures described above we obtain different sets of transport coefficients $\mathrm{R}(T, \psi)$, each of them as a function of equilibrium temperature and chemical potential
difference. Let us use subscript $pc$ for the resistivity matrix obtained from the "perturbation cell" method and $ex$ for the resistivity matrix obtained from the "experiment-like" method. In
each method we calculate the resistivities associated with the gas- and liquid- side measurable heat fluxes using \eqr{eq/TransportCoef/01} and \eqr{eq/TransportCoef/02}.

Furthermore we will use subscript $kin$ for the resistivity matrix obtained from kinetic theory, for which only the gas- side resistivities are available. We calculate the transport coefficients
associated only with mass properties. The corresponding molar coefficients may be calculated using \eqr{eq/TransportCoef/07}. As a result we obtain the following sets of resistivities:
$\mathrm{R}_{pc}^{g}$, $\mathrm{R}_{pc}^{\ell}$, $\mathrm{R}_{ex}^{g}$, $\mathrm{R}_{ex}^{\ell}$, all of which depend on temperature and chemical potential difference as well as on parameters
$\alpha_{qq}$, $\alpha_{1q}$, $\alpha_{11}$. In addition we obtain $\mathrm{R}_{kin}^{g}$ which depend on temperature and chemical potential difference as well as on condensation coefficients
$\sigma_{1}$ and $\sigma_{2}$. We have the following constraints, which they must obey for each $T$ and $\psi$:

- i) the second law consistency;

- ii) the cross coefficients of each $\mathrm{R}$ matrix must satisfy Onsager relations \eqref{eq/TransportCoef/08};

- iii) the corresponding components of $\mathrm{R}_{pc}^{g}$ and $\mathrm{R}_{pc}^{\ell}$ as well as $\mathrm{R}_{ex}^{g}$ and $\mathrm{R}_{ex}^{\ell}$ must satisfy \eqr{eq/TransportCoef/04};

- iv) the corresponding components of $\mathrm{R}_{pc}^{g}$, $\mathrm{R}_{ex}^{g}$ and $\mathrm{R}_{kin}^{g}$ as well as $\mathrm{R}_{pc}^{\ell}$ and $\mathrm{R}_{ex}^{\ell}$ obtained at the
same $T$ and $\psi$ must be equal.

We study the dependence of the different resistivity coefficients on $\alpha_{qq}$, $\alpha_{1q}$ and $\alpha_{11}$ and on $T$ and $\psi $ and their convergence for small $\beta$. We determine
the values of the parameters for which the above constraints are fulfilled.

\subsection{Onsager reciprocal relations.}

In this subsection we investigate the values of parameters $\alpha_{qq}$, $\alpha_{1q}$, $\alpha_{11}$ for which the Onsager relations are fulfilled. This is done for a particular values of
equilibrium temperature and chemical potential difference $T_{eq} = 330$ K and $\psi_{eq} = 700$ J/mol. In \tblsr{tbl/Onsager/Pc-beta-0-0-0}{tbl/Onsager/Ex-beta-0-0-0} we give the relative error
in percent for the gas-side cross coefficients $|(R_{ij}^{g}-R_{ji}^{g})/R_{ij}^{g}|\spd 100\%$ as a function of $\beta $ for $\alpha_{qq}=0$, $\alpha_{1q}=0$, $\alpha_{11}=0$ obtained by
different methods.
\begin{longtable}{l@{\qquad}l@{\qquad}l@{\qquad}l@{\qquad}l@{\qquad}l@{\qquad}l}%
\caption{Relative error in percent for gas-side cross-coefficients obtained by "perturbation cell" method at $T_{eq} = 330$ and $\psi_{eq} = 700$ for different $\beta$ and for $\alpha_{qq} = 0$, $\alpha_{1q} = 0$, $\alpha_{11} = 0$.} \label{tbl/Onsager/Pc-beta-0-0-0}\\%
\hline %
$\beta$  & $R_{q1}$  & $R_{q2}$  & $R_{12}$  \\%
\hline %
2.0e-002     & 8.963066 & 35.863259 & 34.908631 \\%
2.0e-003     & 0.273286 & 0.369082  & 19.683274 \\%
2.0e-004     & 0.011726 & 0.007231  & 1.909391  \\%
2.0e-005     & 0.066375 & 0.071266  & 2.336652  \\%
2.0e-006     & 4.963895 & 8.128243  & 5.843913  \\%
\hline %
\end{longtable} %
\begin{longtable}{l@{\qquad}l@{\qquad}l@{\qquad}l@{\qquad}l@{\qquad}l@{\qquad}l}%
\caption{Relative error in percent for gas-side cross-coefficients obtained by "experiment like" method at $T_{eq} = 330$ and $\psi_{eq} = 700$ for different $\beta$ and for $\alpha_{qq} = 0$, $\alpha_{1q} = 0$, $\alpha_{11} = 0$.} \label{tbl/Onsager/Ex-beta-0-0-0}\\%
\hline %
$\beta$  & $R_{q1}$  & $R_{q2}$  & $R_{12}$  \\%
\hline %
2.0e-002     & 1.275105 & 0.828600  & 754.982200    \\%
2.0e-003     & 0.038759 & 0.363715  & 38.708981 \\%
2.0e-004     & 0.131868 & 0.238584  & 6.247648  \\%
2.0e-005     & 1.301483 & 2.056102  & 20.984734 \\%
2.0e-006     & 13.282959    & 20.788752 & 632.124504    \\%
\hline %
\end{longtable} %
As one can see, $\beta =0.02$ is really an extreme perturbation and the difference is rather large. When we decrease $\beta $ to 2e-4 the differences become small. As we further decrease $\beta$
to 2e-6 the inaccuracy of the numerical solution become comparable to the size of the perturbation. We conclude that the values for $\beta$ to 2e-4 are closest to the converged values and use
them as such.

In \tblsr{tbl/Onsager/Pc-beta-10-10-10}{tbl/Onsager/Ex-beta-10-10-10} we give the same data for the higher continuous resistivities with rather substantial peak, when $\alpha_{qq} = 10$,
$\alpha_{1q} = 10$ and $\alpha_{11} = 10$. As one can see, the Onsager relations are fulfilled there again best for $\beta=$ 2e-4

\begin{longtable}{l@{\qquad}l@{\qquad}l@{\qquad}l@{\qquad}l@{\qquad}l@{\qquad}l}%
\caption{Relative error in percent for gas-side cross-coefficients obtained by "perturbation cell" method at $T_{eq} = 330$ and $\psi_{eq} = 700$ for different $\beta$ and for $\alpha_{qq} = 10$, $\alpha_{1q} = 10$, $\alpha_{11} = 10$.} \label{tbl/Onsager/Pc-beta-10-10-10}\\%
\hline %
$\beta$  & $R_{q1}$  & $R_{q2}$  & $R_{12}$  \\%
\hline %
2.0e-002     & 71.515410    & 78.166809 & 23.572836 \\%
2.0e-003     & 0.745604 & 0.896547  & 0.317348  \\%
2.0e-004     & 0.012358 & 0.012650  & 0.001919  \\%
2.0e-005     & 0.012078 & 0.007485  & 0.005290  \\%
2.0e-006     & 0.713969 & 1.124994  & 0.022121  \\%
\hline %
\end{longtable} %
\begin{longtable}{l@{\qquad}l@{\qquad}l@{\qquad}l@{\qquad}l@{\qquad}l@{\qquad}l}%
\caption{Relative error in percent for gas-side cross-coefficients obtained by "experiment like" method at $T_{eq} = 330$ and $\psi_{eq} = 700$ for different $\beta$ and for $\alpha_{qq} = 10$, $\alpha_{1q} = 10$, $\alpha_{11} = 10$.} \label{tbl/Onsager/Ex-beta-10-10-10}\\%
\hline %
$\beta$  & $R_{q1}$  & $R_{q2}$  & $R_{12}$  \\%
\hline %
2.0e-002     & 4.225362 & 2.559393  & 12.259260 \\%
2.0e-003     & 0.443944 & 0.256804  & 1.091842  \\%
2.0e-004     & 0.068621 & 0.019788  & 0.093041  \\%
2.0e-005     & 0.269764 & 0.407090  & 0.008844  \\%
2.0e-006     & 2.717575 & 4.149484  & 2.025054  \\%
\hline %
\end{longtable} %

The similar picture is observed for the liquid-side resistivities and we do not give those data here.

We may notice that the behavior of the resistivities with respect to $\beta$ is independent on the behavior of the resistivities with respect to $\alpha_{qq}$, $\alpha_{1q}$ and $\alpha_{11}$.
This is natural, as these parameters control the different aspects of the system: $\beta$ controls the perturbation rate, while $\alpha$'s are adjustable parameters, which control the size of
the peak in the continuous resistivities.

\subsection{Second law consistency.}

In this subsection we investigate the values of parameters $\alpha_{qq}$, $\alpha_{1q}$, $\alpha_{11}$ for which the second law of thermodynamics are fulfilled. That is that the excess entropy
production is positive and therefore the matrix of the resistivity coefficients is positive definite. This requires that the diagonal coefficients are positive and for each pair $q1$, $q2$ and
$12$ of the cross coefficients the expression
\begin{equation}\label{eq/Results/01}
DR_{ik} \equiv R_{ii}R_{kk} - {1 \over 4}(R_{ik}+R_{ki})^{2} > 0
\end{equation}
must be positive.

In \tblr{tbl/2ndLaw/Pc-4-Aqq-0-0} we give the diagonal coefficients and expression \eqref{eq/Results/01} for each pair of the cross coefficients as a function of $\alpha_{qq}$ for $\alpha_{1q} =
0$, $\alpha_{11} = 0$ and $\beta =$ 2e-4 obtained by the "perturbation cell" method. In \tblsr{tbl/2ndLaw/Pc-4-0-Aiq-0}{tbl/2ndLaw/Pc-4-0-0-Aii} we give the same quantities for other choices of
$\alpha$.
\begin{longtable}{l@{\qquad}l@{\qquad}l@{\qquad}l@{\qquad}l@{\qquad}l@{\qquad}l}%
\caption{2nd law consistency for gas-side coefficients. The diagonal coefficients and the quantities defined by \eqref{eq/Results/01}. Data are obtained by "perturbation cell" method at $T_{eq} = 330$ and $\psi_{eq} = 700$ for different $\alpha_{qq}$ and for $\beta = 0.0002$, $\alpha_{1q} = 0$, $\alpha_{11} = 0$.} \label{tbl/2ndLaw/Pc-4-Aqq-0-0}\\%
\hline %
$\alpha_{qq}$    & $R_{qq}$  & $R_{11}$  & $R_{22}$  & $DR_{q1}$  & $DR_{q2}$  & $DR_{12}$  \\%
\hline %
0    & 7.05644e-015 & 0.0754717 & -0.0919278    & 2.13025e-015  & -2.59473e-015 & -0.0277518    \\%
1    & 3.36047e-012 & 0.0937784 & -0.0741586    & 1.26056e-012  & -9.9683e-013  & -0.0278179    \\%
10   & 3.35408e-011 & 0.259425  & 0.0851534 & 3.48053e-011  & 1.14244e-011  & 0.0874467 \\%
\hline %
\end{longtable} %
\begin{longtable}{l@{\qquad}l@{\qquad}l@{\qquad}l@{\qquad}l@{\qquad}l@{\qquad}l}%
\caption{2nd law consistency for gas-side coefficients. The diagonal coefficients and the quantities defined by \eqref{eq/Results/01}. Data are obtained by "perturbation cell" method at $T_{eq} = 330$ and $\psi_{eq} = 700$ for different $\alpha_{1q}$ and for $\beta = 0.0002$, $\alpha_{qq} = 0$, $\alpha_{11} = 0$.} \label{tbl/2ndLaw/Pc-4-0-Aiq-0}\\%
\hline %
$\alpha_{1q}$    & $R_{qq}$  & $R_{11}$  & $R_{22}$  & $DR_{q1}$  & $DR_{q2}$  & $DR_{12}$  \\%
\hline %
0    & 7.05644e-015 & 0.0754717 & -0.0919278    & 2.13025e-015  & -2.59473e-015 & -0.0277518    \\%
1    & 7.05608e-015 & 0.0746391 & -0.0910331    & 2.10664e-015  & -2.56935e-015 & -0.0271785    \\%
10   & 7.05304e-015 & 0.0670813 & -0.0828915    & 1.89251e-015  & -2.33855e-015 & -0.0222419    \\%
\hline %
\end{longtable} %
\begin{longtable}{l@{\qquad}l@{\qquad}l@{\qquad}l@{\qquad}l@{\qquad}l@{\qquad}l}%
\caption{2nd law consistency for gas-side coefficients. The diagonal coefficients and the quantities defined by \eqref{eq/Results/01}. Data are obtained by "perturbation cell" method at $T_{eq} = 330$ and $\psi_{eq} = 700$ for different $\alpha_{11}$ and for $\beta = 0.0002$, $\alpha_{qq} = 0$, $\alpha_{1q} = 0$.} \label{tbl/2ndLaw/Pc-4-0-0-Aii}\\%
\hline %
$\alpha_{11}$    & $R_{qq}$  & $R_{11}$  & $R_{22}$  & $DR_{q1}$  & $DR_{q2}$  & $DR_{12}$  \\%
\hline %
0    & 7.05644e-015 & 0.0754717 & -0.0919278    & 2.13025e-015  & -2.59473e-015 & -0.0277518    \\%
1    & 7.05717e-015 & 0.370078  & 0.265626  & 1.04468e-014  & 7.49827e-015  & 0.381226  \\%
10   & 7.10378e-015 & 3.02063   & 3.48284   & 8.58316e-014  & 9.89654e-014  & -69.2846  \\%
\hline %
\end{longtable} %

We see, that the required quantities become positive for rather big values of $\alpha_{qq}$. They almost do not depend on the value of $\alpha_{1q}$ and they are positive for moderate values of
parameter $\alpha_{11}$. It is clear that finite values of $\alpha_{qq}$ and $\alpha_{11}$ are needed to have a positive excess entropy production.

All the above quantities almost do not depend on the value of $\beta$ in the range [1e-5, 1e-3]. The "experimental-like" procedure leads to almost the same values of all the quantities. The
liquid-side coefficients reveal a similar behavior.

\subsection{Gas- and liquid- coefficients.}

In this subsection we investigate the validity of \eqr{eq/TransportCoef/04}. In \tblr{tbl/Bulk/Pc-4-1-1-1} we give the relative error in percent between the left hand side and the right hand
side of \eqr{eq/TransportCoef/04}.
\begin{longtable}{l@{\qquad}l@{\qquad}l@{\qquad}l@{\qquad}l@{\qquad}l@{\qquad}l@{\qquad}l@{\qquad}l@{\qquad}l}%
\caption{Relative error in percent for invariant expressions in \eqr{eq/TransportCoef/04} obtained by "perturbation cell" method at $T_{eq} = 330$ and $\psi_{eq} = 700$ for $\beta = 0.0002$ and $\alpha_{qq} = 1$, $\alpha_{1q} = 1$, $\alpha_{11} = 1$.} \label{tbl/Bulk/Pc-4-1-1-1}\\%
\hline %
${qq}$     & ${11}$  & ${22}$  & ${q1}$  & ${1q}$  & ${q2}$  & ${2q}$  & ${12}$  & ${21}$   \\%
\hline %
 0.000000   & 0.000002  & 0.000085  & 0.000001  & 0.000389  & 0.000001  & 0.000389  & 0.000060  & 0.000003  \\%
\hline %
\end{longtable} %
For instance, the ${q1}$ quantity is equal to $|(R_{q1}^{\ell} - {h}_{1,\,eq}^{\ell}\,R_{qq}^{\ell}) - (R_{q1}^{g} - {h}_{1,\,eq}^{g}\,R_{qq}^{g})|/|R_{q1}^{\ell} -
{h}_{1,\,eq}^{\ell}\,R_{qq}^{\ell}|\spd100\%$. The other quantities are defined in the same way. These errors almost do not depend neither on the value of $\beta$ in the range [1e-5, 1e-3] nor
on the values of $\alpha_{qq}$, $\alpha_{1q}$, $\alpha_{11}$. The "experimental-like" procedure leads to almost the same results.

\subsection{Comparison to kinetic theory.}

In this subsection we investigate the values of parameters  $\alpha_{qq}$, $\alpha_{1q}$, $\alpha_{11}$ which makes the coefficients agree with the kinetic theory coefficients. We do it for
$\beta$ = 2e-4 as this perturbation rate gives the most accurate results. We again do this for temperature $T_{eq} = 330$ K and chemical potential difference $\psi_{eq} = 700$ J/mol. The values
of parameters, used for kinetic theory are the same, as we use in our calculations. Particularly, the heat conductivities are $\lambda_{1} = $0.0140 W/(m K) and $\lambda_{2} = $0.0157 W/(m K),
$M_{1} = 84.162$  g/mol and $M_{2} = 86.178$ g/mol. We compare here only the "perturbation cell" method with kinetic theory.

We found that the variation of $\alpha_{1q}$ from 0 to 10 makes the diagonal coefficients vary about 1\% and the cross coefficients vary not more then 5\%. As the variation of $\alpha_{1q}$ is
quite substantial, the variation in the coefficients which it induces is negligible. We therefore take $\alpha_{1q} = 0$ in all further analysis.

For the above parameters $R_{qq,\,kin} = 2.96792 \times 10^{-11}$. We found that $R_{qq,\,pc}$ is practically independent on $\alpha_{11}$ while it depends linearly on $\alpha_{qq}$, see
\figr{R_{qq}-A_{qq}}. One can see from the plot, that $R_{qq,\,kin} = R_{qq,\,pc}$ for $\alpha_{qq} \approx 9$.
\begin{figure}[hbt!]
\centering
\includegraphics[scale=\profilescale]{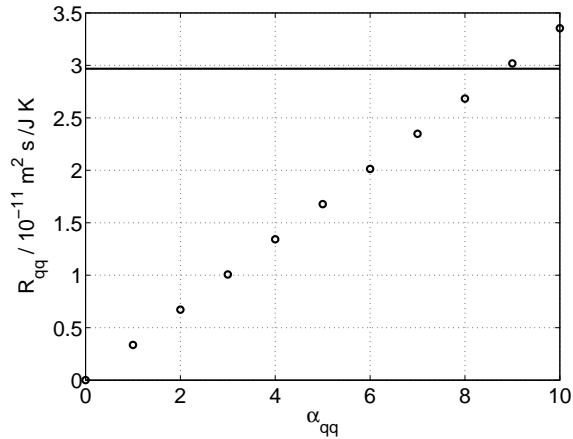}
\caption{Dependence of $R_{qq}$ on $\alpha_{qq}$ obtained by "perturbation cell" method at $T_{eq} = 330$ and $\psi_{eq} = 700$ for $\alpha_{1q} = 0$ and $\alpha_{11} = 1$. $R_{qq,\,kin}$ is
drawn as a constant line.}\label{R_{qq}-A_{qq}}
\end{figure}

The diagonal coefficients $R_{11,\,pc}$ and $R_{22,\,pc}$ depend both on $\alpha _{qq}$ and $\alpha _{11}$. Since we have found the corresponding to kinetic theory value of $\alpha _{qq}$ by
mapping the $R_{qq}$ coefficient, we will further investigate the dependence of $R_{11,\,pc}$ and $R_{22,\,pc}$ using this value of $\alpha_{qq}$ and varying only $\alpha_{11}$. The diagonal
coefficients $R_{11,\,kin}$ and $R_{22,\,kin}$ depend, in their turn, on the condensation coefficients $\sigma_{1}$ and $\sigma_{2}$ respectively. We plot this dependence in the same plot with
the dependency of $R_{ii,\,pc}$ on $\alpha _{11}$, see \figr{R_{ii}-A_{11}}. The dependence of $R_{ii,\,pc}$ on $\alpha_{11}$ is given by the dotted line with the values of $\alpha_{11}$ drawn
on the bottom $x$-axes. The dependence of $R_{ii,\,kin}$ on $\sigma_{i}$ is given by the solid line with the values $\sigma_{i}$ drawn on the top $x$-axes.
\begin{figure}
\centering
\subfigure
{\includegraphics[scale=\profilescale]{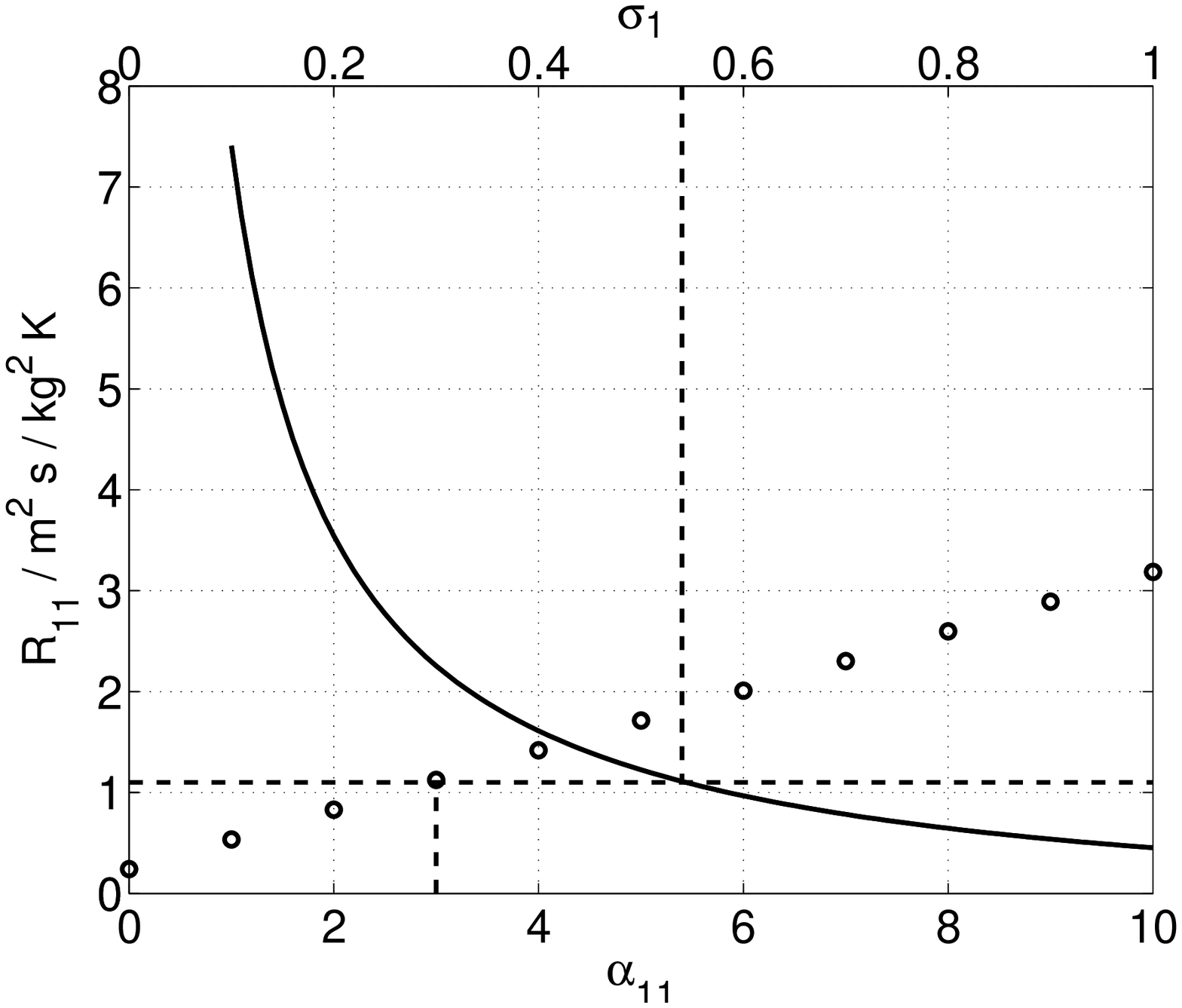}\label{R_{11}-A_{11},S_{1}} } %
\subfigure
{\includegraphics[scale=\profilescale]{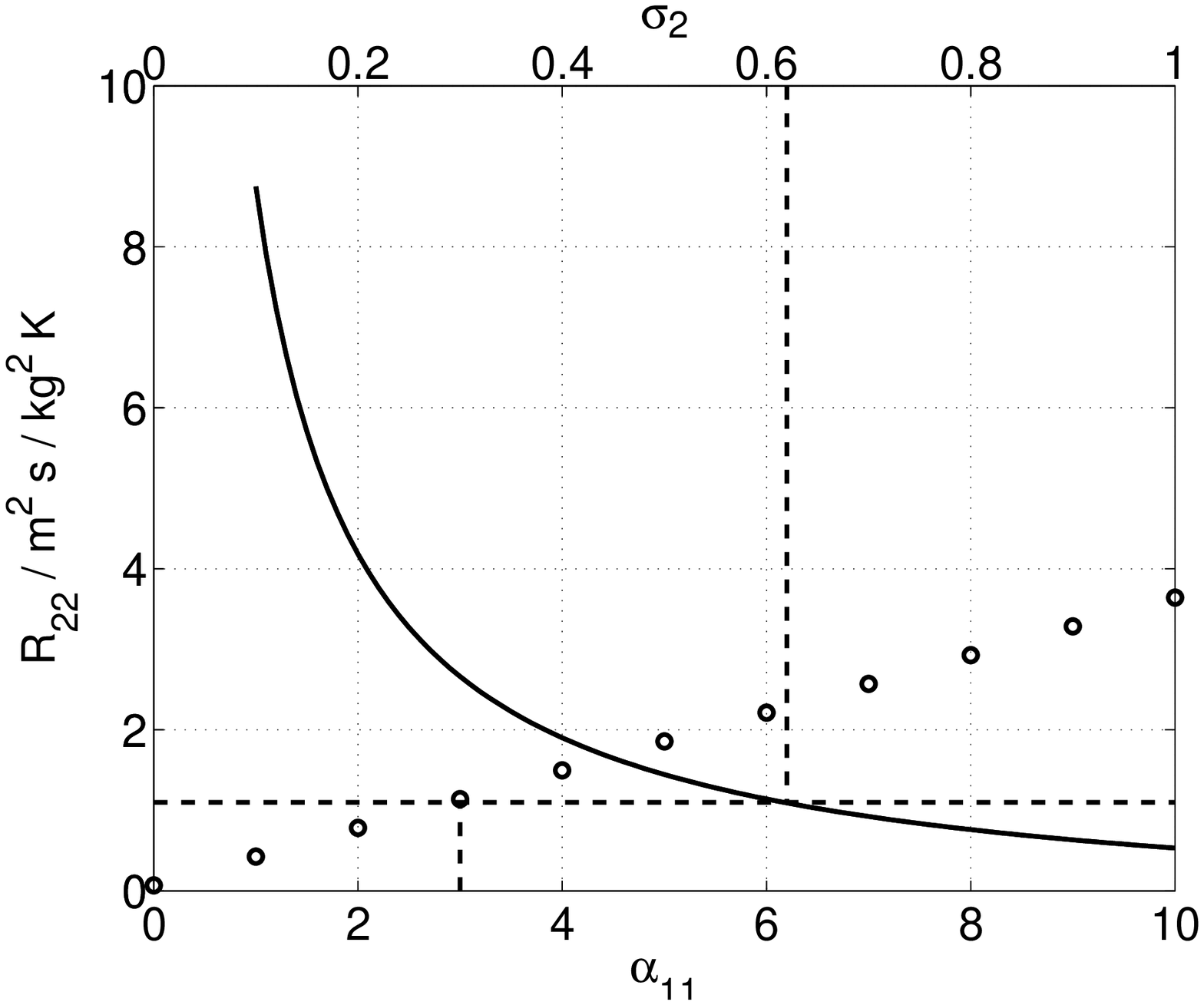} \label{R_{22}-A_{11,S_{2}}} } %
\caption{Dependence of $R_{11,\,pc}$ and $R_{22,\,pc}$ on $\alpha_{11}$ (dots, bottom axes) and $R_{11,\,kin}$ and $R_{22,\,kin}$ on $\sigma_{1}$ and $\sigma_{2}$ (curve, top axes),
respectively. Data are obtained at $T_{eq} = 330$ and $\psi_{eq} = 700$ for $\alpha_{qq} = 9$ and $\alpha_{1q} = 0$.}\label{R_{ii}-A_{11}}
\end{figure}

Consider a particular value $R_{ii,\,0}$ of the diagonal coefficient $R_{ii}$, where $i$ is either 1 or 2, which is indicated by a horizontal dashed line on a figure. To find the value of
$\alpha_{11}$ for which $R_{ii,\,pc} = R_{ii,\,0}$ we draw a perpendicular from the point where it crosses the dotted line to the bottom axes. To find the value of $\sigma_{i}$ for which
$R_{ii,\,kin} = R_{ii,\,0}$ we draw a perpendicular from the point where the horizontal dashed line crosses the solid line to the top axes. For instance, the value $R_{22,\,0} = 1.1$ corresponds
to $\alpha_{11} = 3$ and $\sigma_{2} = 0.62$. The value $\alpha_{11} = 3$, in its turn, gives $R_{11,\,0} = 1.1$ which corresponds to $\sigma_{1} = 0.54$.

One may start by specifying $\alpha_{11}$, rather then $R_{ii,\,0}$, to find $\sigma_{1}$ and $\sigma_{2}$. Then we draw a perpendicular from the bottom axes until it crosses the dotted line,
which gives the value $R_{ii,\,0}$ of $R_{ii,\,pc}$. Given the value of $R_{ii,\,kin}$ to be the same, we find the value of $\sigma_{i}$ as described above. For the above example $\alpha_{11} =
3$ corresponds to $\sigma_{1} = 0.54$ and $\sigma_{2} = 0.62$. We see, that we may not specify both $\sigma_{1}$ and $\sigma_{2}$ independently: they must have the values which both correspond
to the same $\alpha_{11}$. For similar components, like those we are interested in, $\sigma_{1}$ and $\sigma_{2}$ should not differ much from each other, and therefore $\alpha_{11}$, a
coefficient which is related to the diffusion of one component through the other, should reflect this difference.


Having the diagonal coefficient mapped we have the parameters $\alpha_{qq}$ and $\alpha_{11}$ defined uniquely (and taking into account that $\alpha_{1q}$ has negligible effect), as well as
$\sigma_{1}$ and $\sigma_{2}$ for kinetic theory. We now compare the values of the cross coefficients given by "perturbation cell" method and kinetic theory.
\begin{longtable}{l@{\qquad}l@{\qquad}l@{\qquad}l@{\qquad}l@{\qquad}l@{\qquad}l} %
\caption{Gas-side transport coefficients obtained from kinetic theory and by "perturbation cell" method at $T_{eq} = 330$ and $\psi_{eq} = 700$ for $\beta = 0.0002$.}\label{tbl/Coeff/Kn-Pc-4-9-0-3}\\%
\hline %
parameters   & $R_{qq}$  & $R_{11}$  & $R_{22}$  & $R_{q1}$  & $R_{q2}$  & $R_{12}$  \\%
\hline %
\begin{tabular}{l} $\sigma_{1}=0.54$ \\ $\sigma_{2}=0.62$ \end{tabular} & 2.96792e-011   & 1.11091   & 1.09136   & 3.82826e-007     & 4.41483e-007     & 0.0130511      \\%
\begin{tabular}{l} $\alpha_{qq}=9$ \\ $\alpha_{1q}=0$ \\ $\alpha_{11}=3$\end{tabular} & 3.01874e-011     & 1.12461   & 1.13991   & 2.31477e-006     & 2.27003e-006     & -0.816559      \\%
\hline %
\end{longtable} %
One can see from \tblr{tbl/Coeff/Kn-Pc-4-9-0-3} that while the diagonal coefficients are the same\footnote{One should not expect exact compatibility between kinetic theory, which is most
appropriate for gases with short range potentials, and the gradient theory, which is most appropriate for fluids with long range potentials. The purpose of this comparison in not to determine
the exact values of adjustable parameters, but to show that it is possible to match coefficients in the two theories and to show the typical values of the parameters.}, the cross coefficients we
find are between 1-2 orders of magnitude larger than those found by kinetic theory. $R_{12}$ even has a different sign.

\subsection{Temperature and chemical potential difference dependence.}

\begin{figure}[hbt!]
\centering
\includegraphics[scale=\profilescale]{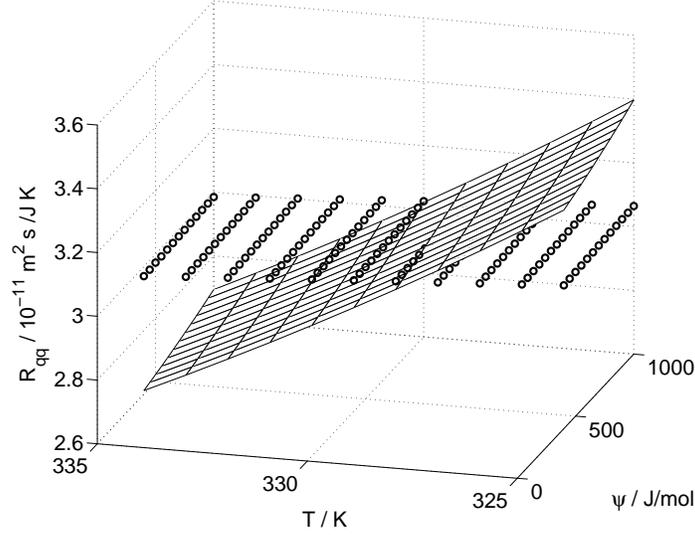}
\caption{Dependence of $R_{qq}$ on $T$ and $\psi$ obtained from kinetic theory for $\sigma_{1}=0.54$ and $\sigma_{2}=0.62$ (plane) and by "perturbation cell" method for $\alpha_{qq} = 9$,
$\alpha_{1q} = 0$ and $\alpha_{11} = 3$ (points).}\label{Rqq}
\end{figure}

In this subsection we investigate the dependence of the resistivity coefficients on the temperature and the chemical potential difference. On \figsr{Rqq}{R11} we plot the these dependencies for
$R_{qq}$, $R_{q1}$ and $R_{11}$ coefficients obtained from kinetic theory and "perturbation cell" method for the range of temperatures $[325,\ldots,335]$ and for the range of chemical potential
differences $[400,\ldots,1000]$.
\begin{figure}[hbt!]
\centering
\includegraphics[scale=\profilescale]{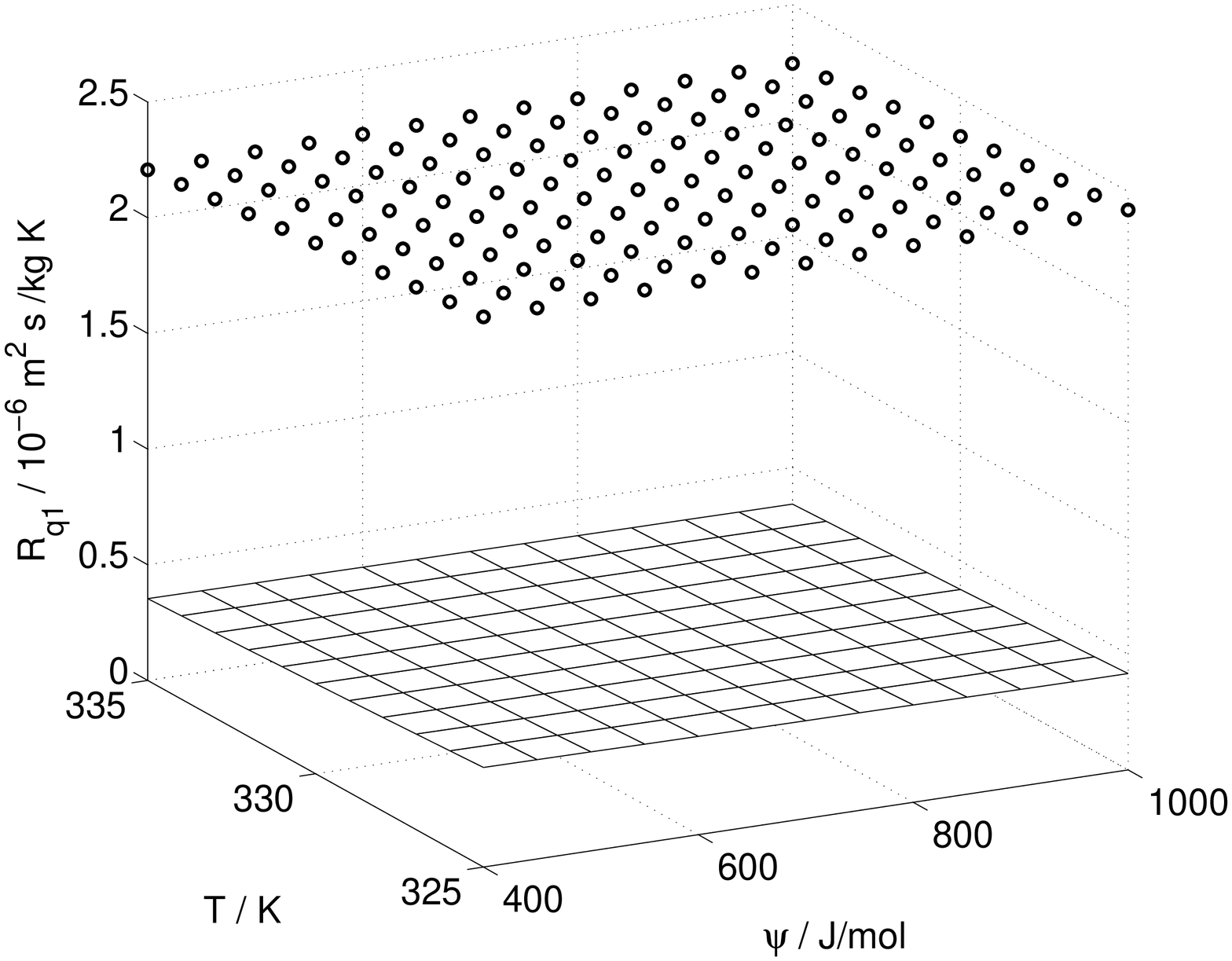}
\caption{Dependence of $R_{q1}$ on $T$ and $\psi$ obtained from kinetic theory for $\sigma_{1}=0.54$ and $\sigma_{2}=0.62$ (plane) and by "perturbation cell" method for $\alpha_{qq} = 9$,
$\alpha_{1q} = 0$ and $\alpha_{11} = 3$ (points).}\label{Rq1}
\end{figure}
\begin{figure}[hbt!]
\centering
\includegraphics[scale=\profilescale]{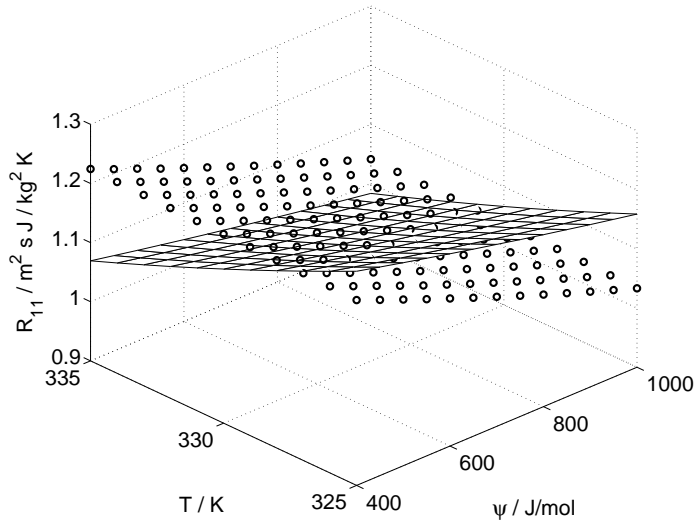}
\caption{Dependence of $R_{11}$ on $T$ and $\psi$ obtained from kinetic theory for $\sigma_{1}=0.54$ and $\sigma_{2}=0.62$ (plane) and by "perturbation cell" method for $\alpha_{qq} = 9$,
$\alpha_{1q} = 0$ and $\alpha_{11} = 3$ (points).}\label{R11}
\end{figure}

The domain of $T$ and $\psi$ is not big, so the dependence on them is linear, as expected.

\section{Discussion and conclusions.}\label{sec/Discussion}

In this paper we have studied stationary transport of heat and mass through the liquid-vapor interface in a mixture, using the square gradient theory \cite{glav/grad1, glav/grad2}. We derived an
expression for the excess entropy production of a surface from the continuous description, which is identical to the one derived directly for the discrete description using the property of local
equilibrium \cite{kjelstrupbedeaux/heterogeneous}. This makes it possible to give the linear force-flux relations for this case. These relations involve the interfacial resistivities or transfer
coefficients, which were the main focus of interest in this paper. Given the numerical solutions of the non-equilibrium gradient model we were able to calculate these coefficients directly for a
two-component mixture. This gives an independent way to determine the interfacial resistivities.

The main input parameters of the model, aside from the parameters in the equilibrium square gradient model, are the local resistivity profiles used to calculate the continuous solution. There is
not much theoretical information about the numerical value of these resistivities. In the vapor phase one can use kinetic theory. In the liquid phase it is most appropriate to use experimental
values. There is no experimental information about the local resistivities in the interfacial region. As the local resistivities change in the surface from one bulk value to the other, it is
natural to assume that they contain a contribution similar to the profile of the order parameter. There is also evidence from molecular dynamics simulations for one-component systems
\cite{surfres} that there is a peak in the local resistivities in the surface. As we are in the framework of the gradient theory, it is naturally to assume that this peak is caused by a square
gradient term, which is similar to the gradient contribution, which the Helmholtz energy density has in the interfacial region, namely $|\nabla\rho|^{2}$. The amplitude of this peak is not given
by any theory and is used as a parameter. We therefore get that each of three local resistivities for a two-component mixture has the form given in \eqr{eq/Equations/Phenomenological/03}. Thus
we get three adjustable amplitudes, $\alpha_{qq}$, $\alpha_{1q}$ and $\alpha_{11}$, two of which are found to contribute significantly to the value of the transfer coefficients.

In order to determine the typical values of the $\alpha $'s we need to compare our results with independently obtained resistivities. Unfortunately, not much experimental data are available for
multi-component resistivities and, to the best of our knowledge, no data are available for our system. Furthermore, no molecular dynamic simulations are available for mixtures. The only
reasonable source of comparison is kinetic theory, which gives the expressions for the interfacial resistivities or transfer coefficients \cite{Cipolla1974, Bedeaux1990}. We therefore compare
our results to kinetic theory. Having three adjustable parameters in the gradient theory, $\alpha_{qq}$, $\alpha_{1q}$ and $\alpha_{11}$, and two adjustable parameters in kinetic theory, the
condensation parameters $\sigma_{1}$ and $\sigma_{2}$, we are able to match three diagonal coefficients $R_{qq}$, $R_{11}$ and $R_{22}$. We found that $R_{qq}$ does not really depend on
$\alpha_{1q}$ and $\alpha_{11}$. This makes it possible to fit $\alpha_{qq}$ using $R_{qq}$ alone. For the values of the temperature and chemical potentials considered this gave $\alpha_{qq}
\simeq 9$. We furthermore found that the interfacial resistivities did not really depend on $\alpha_{1q}$. We therefore took this amplitude equal to zero. In kinetic theory $R_{11}$ and $R_{22}$
depend on the condensation coefficients $\sigma_{1}$ and $\sigma_{2}$, respectively. Choosing $\alpha_{11}=3$ gives values for the condensation coefficients of 0.54 and 0.62. As the components
considered are very similar it is to be expected that these coefficients are close to each other. The values of $\alpha$'s obtained from the matching are such that the excess entropy production
of the surface is positive, the second law is obeyed and the Onsager relations are valid. Having found the values of the $\alpha$'s from the diagonal transfer coefficients the value of the cross
coefficients follows uniquely.

We found that the values of the cross coefficients, obtained by our method are between 1-2 orders of magnitude larger than those found from kinetic theory. This confirms results from molecular
dynamics simulations \cite{jialin/longrange}, where it was found that increasing the range of the attractive potential increased in particular the cross coefficients substantially above the
values predicted by kinetic theory. This is an interesting result, indicating that kinetic theory underestimates the transfer coefficients for real fluids. This also indicates, that the effect
of coupling will be important in the interfacial region. Experiments also confirm the importance of the cross coefficients \cite{Mills2002, James2006}.

The effect of cross coefficients can be related to the measurable quantities, such as measurable heat of transfer $q^{*}_{i} \equiv - R_{qi}/R_{qq}$. This quantity can be associated both with
gas and liquid phases in accordance to the corresponding heat fluxes. The difference $q^{*,g}_{i} - q^{*,\ell}_{i} = - (R_{qi}^{g}-R_{qi}^{\ell})/R_{qq} = - (h_{i,eq}^{g}-h_{i,eq}^{\ell})$ is
equal to the difference of partial enthalpies between gas and liquid in equilibrium. This quantity is substantial, which indicates that $q^{*,g}_{i} - q^{*,\ell}_{i}$ is also substantial. This,
in turn, makes the difference between the cross coefficients on gas and liquid side to be non-vanishing. This gives a theoretical ground for the importance of coupling in the interfacial region.
Experiments \cite{Mills2002, James2006} confirm the size and importance of the heat of transfer on the gas side.

We did the comparison for one value of the temperature and chemical potential only. If one extends the analysis to a larger domain, one finds that the $\alpha$'s depend on the temperature and
the chemical potential difference; we refer to \cite{bedeaux/vdW/III} in this context. The results of kinetic theory \cite{Pao1971a, Sone1973, Cipolla1974, Bedeaux1990} and molecular dynamics
\cite{Simon2004} both support the existence of a peak in the diagonal local resistivities and therefore the use of finite values for $\alpha_{qq}$ and $\alpha_{11}$.

\begin{acknowledgements}
We want to thank Eivind Johannessen for advise. We are also grateful to NFR for Storforks grant no.167336/V30.
\end{acknowledgements}

\appendix

\section{Excess in curvilinear coordinates.}\label{sec/Appendix/Excess}
\subsection{On the definition of an excess quantity using curvilinear coordinates.}\label{sec/Appendix/Excess/Def}

One may think of an alternative definition of an excess quantity
\begin{equation}\label{eq/Appendix/Excess/Def/01}
\widehat{\phi}_{a}(\xs, x_{2}, x_{3}) \equiv \int_{\displaystyle \xgsb}^{\displaystyle \xlsb}{dx_{1}\,h_{1}\,\phi^{ex}(\vR; \xs)}
\end{equation}
We note however, that $\widehat{\phi}_{a}$  has no physical meaning, while $\widehat{\phi}$ has. The reason for this is that
\begin{equation}\label{eq/Appendix/Excess/Def/02}
\Phi \equiv \int\!\!\!\!\int_{S}{dx_{2}\,dx_{3}\,h_{2}^{s}\,h_{3}^{s}\,\widehat{\phi}(\xs, x_{2}, x_{3})} = \int\!\!\!\!\int_{S}\int_{\displaystyle \xgsb}^{\displaystyle
\xlsb}{dx_{1}\,dx_{2}\,dx_{3}\,h_{1}\,h_{2}\,h_{3}\,\phi^{ex}(\vR; \xs)}
\end{equation}
is the total amount of some physical quantity in the volume which is limited by the surfaces $S$ at $\xgsb$ and $\xlsb$, while
\begin{equation}\label{eq/Appendix/Excess/Def/03}
\Phi_{a} \equiv \int\!\!\!\!\int_{S}{dx_{2}\,dx_{3}\,h_{2}^{s}\,h_{3}^{s}\,\widehat{\phi}_{a}(\xs, x_{2}, x_{3})} = \int\!\!\!\!\int_{S}\int_{\displaystyle \xgsb}^{\displaystyle
\xlsb}{dx_{1}\,dx_{2}\,dx_{3}\,h_{1}\,h_{2}^{s}\,h_{3}^{s}\,\phi^{ex}(\vR; \xs)}
\end{equation}
is not. If the interfacial thickness is small compared to the radii of curvature, the difference between $\widehat{\phi}$ and $\widehat{\phi}_{a}$ is small and vanishes for planar interface
considered under cartesian coordinates. However, it is $\Phi$ but not $\Phi_{a}$ which is a physical amount, and thus $\widehat{\phi}$ is the surface density.

\subsection{On the integration of a gradient function in curvilinear coordinates.}\label{sec/Appendix/Excess/Gradient}

\paragraph{}

Consider a function $\phi$ being the divergence of a vector function: $\phi = \nabla\spd\vq(\vR)$. Then
\begin{equation}\label{eq/Appendix/Excess/Gradient/01}
\begin{array}{rl}
(\nabla\spd\vq)^{ex}(\vR; \xs) &= \nabla\spd\vq(\vR) - \{\nabla\spd\vq^{g}(\vR)\}\,\Theta(\xs-x_{1}) - \{\nabla\spd\vq^{\ell}(\vR)\}\,\Theta(x_{1}-\xs) \\\\
&= \nabla\spd(\vq^{ex})(\vR; \xs) + \vq^{g}(\vR)\spd\nabla\Theta(\xs-x_{1}) + \vq^{\ell}(\vR)\spd\nabla\Theta(x_{1}-\xs)
\end{array}
\end{equation}
where
\begin{equation}\label{eq/Appendix/Excess/Gradient/02}
\vq^{ex}(\vR; \xs) \equiv \vq(\vR) - \vq^{g}(\vR)\,\Theta(\xs-x_{1}) - \vq^{\ell}(\vR)\,\Theta(x_{1}-\xs)
\end{equation}
and $\vq^{g}$ and $\vq^{\ell}$ are defined similarly to $\phi^{g}$ and $\phi^{\ell}$. Furthermore its excess
\begin{equation}\label{eq/Appendix/Excess/Gradient/03}
\widehat{\nabla\spd\vq}(\xs, \vR_{\parallel}) = \displaystyle \frac{1}{h_{2}^{s}\,h_{3}^{s}}\,\int_{\displaystyle \xgsb}^{\displaystyle
\xlsb}{dx_{1}\,h_{1}\,h_{2}\,h_{3}\,(\nabla\spd\vq)^{ex}(\vR; \xs)}
\end{equation}

Using the standard formula for the divergence of a vectorial function in curvilinear coordinates
\begin{equation}\label{eq/Appendix/Excess/Gradient/04}
\nabla\spd\vq = \displaystyle \frac{1}{h_{1}\,h_{2}\,h_{3}}\left(\frac{\partial}{\partial x_{1}}(h_{2}h_{3}\,q_{1}) + \frac{\partial}{\partial x_{2}}(h_{1}h_{3}\,q_{2}) +
\frac{\partial}{\partial x_{3}}(h_{1}h_{2}\,q_{3}) \right)
\end{equation}
one can show that
\begin{equation}\label{eq/Appendix/Excess/Gradient/05}
\begin{array}{rl}
&\displaystyle \int_{\displaystyle \xgsb}^{\displaystyle \xlsb}{dx_{1}\,h_{1}\,h_{2}\,h_{3}\,\nabla\spd(\vq^{ex})(\vR; \xs)} %
= \\\\ = %
&\displaystyle \left.h_{2}h_{3}\,q_{\perp}^{ex}\vphantom{\int}\right|_{\displaystyle\xgsb}^{\displaystyle\xlsb} + \int_{\displaystyle \xgsb}^{\displaystyle
\xlsb}{dx_{1}\,\left(\frac{\partial}{\partial x_{2}}(h_{1}h_{3}\,q_{2}^{ex}) + \frac{\partial}{\partial x_{3}}(h_{1}h_{2}\,q_{3}^{ex}) \right)}
\end{array}
\end{equation}

Using the standard formula for the gradient of a scalar function in curvilinear coordinates
\begin{equation}\label{eq/Appendix/Excess/Gradient/06}
\nabla\theta =  \displaystyle \frac{1}{h_{1}}\frac{\partial \theta}{\partial x_{1}}\,\mathbf{i_{1}} + \frac{1}{h_{2}}\frac{\partial \theta}{\partial x_{2}}\,\mathbf{i_{2}} +
\frac{1}{h_{3}}\frac{\partial \theta}{\partial x_{3}}\,\mathbf{i_{3}}
\end{equation}
one can show that for Heaviside step function $\Theta$
\begin{equation}\label{eq/Appendix/Excess/Gradient/07}
\displaystyle \frac{1}{h_{2}^{s}\,h_{3}^{s}}\,\int_{\displaystyle\xgsb}^{\displaystyle\xlsb}{dx_{1}\,h_{1}\,h_{2}\,h_{3}\,\vq^{b}(\vR)\spd\nabla\Theta(x_{1}-\xs)} = \vq^{b}(\xs,
\vR_{\parallel})\spd\mathbf{i_{1}} \equiv q_{\perp}^{b}(\xs, \vR_{\parallel})
\end{equation}

Substituting \eqr{eq/Appendix/Excess/Gradient/01} into \eqr{eq/Appendix/Excess/Gradient/03} and using \eqr{eq/Appendix/Excess/Gradient/05} and \eqr{eq/Appendix/Excess/Gradient/07} we obtain
\begin{equation}\label{eq/Appendix/Excess/Gradient/09}
\widehat{\nabla\spd\vq}(\xs, \vR_{\parallel}) = q_{\perp}^{\ell}(\xs, \vR_{\parallel}) - q_{\perp}^{g}(\xs, \vR_{\parallel}) + \widehat{\nabla_{\parallel}\spd\vq_{\parallel}}(\xs,
\vR_{\parallel})
\end{equation}
where
\begin{equation}\label{eq/Appendix/Excess/Gradient/10}
\widehat{\nabla_{\parallel}\spd\vq_{\parallel}} = \frac{1}{h_{2}^{s}\,h_{3}^{s}}\,\int_{\displaystyle \xgsb}^{\displaystyle \xlsb}{dx_{1}\left(\frac{\partial}{\partial
x_{2}}(h_{1}h_{3}\,q_{2}^{ex}) + \frac{\partial}{\partial x_{3}}(h_{1}h_{2}\,q_{3}^{ex}) \right)}
\end{equation}
and we took into account that according to \eqr{eq/Excess/Definition/01} $q_{\perp}^{ex}(\xgsb) = q_{\perp}^{ex}(\xlsb) = 0$.

\paragraph{}

Consider a special form of a vector $\vq$ for which $\vq_{\parallel} = \vJ_{\parallel}\phi$ where $\nabla_{\parallel}\spd\vJ_{\parallel} = 0$. Here $\nabla_{\parallel}$ is a parallel component
of three-dimensional nabla-operator so that
\begin{equation}\label{eq/Appendix/Excess/Gradient/12}
\begin{array}{rl}
\displaystyle \nabla_{\parallel}\spd\vq_{\parallel} &= \displaystyle \frac{1}{h_{1}\,h_{2}\,h_{3}}\left(\frac{\partial}{\partial x_{2}}(h_{1}h_{3}\,q_{2}) + \frac{\partial}{\partial
x_{3}}(h_{1}h_{2}\,q_{3}) \right)
\\\\
\displaystyle \nabla_{\parallel}\theta &= \displaystyle \frac{1}{h_{2}}\frac{\partial \theta}{\partial x_{2}}\,\mathbf{i_{2}} + \frac{1}{h_{3}}\frac{\partial \theta}{\partial
x_{3}}\,\mathbf{i_{3}}
\end{array}
\end{equation}
Then
\begin{equation}\label{eq/Appendix/Excess/Gradient/11}
\displaystyle \frac{\partial}{\partial x_{2}}(h_{1}h_{3}\,q_{2}) + \frac{\partial}{\partial x_{3}}(h_{1}h_{2}\,q_{3}) = h_{1}h_{2}h_{3}\left(\vJ_{\parallel}\spd(\nabla_{\parallel}\phi) +
(\nabla_{\parallel}\spd\vJ_{\parallel})\phi \right) = h_{1}h_{2}h_{3}\,\vJ_{\parallel}\spd(\nabla_{\parallel}\phi)
\end{equation}
%

\bibliographystyle{unsrt}

\end{document}